\documentclass[11pt]{article}

\usepackage[margin=1in]{geometry}
\usepackage{graphicx}
\usepackage{array}
\usepackage{float}
\usepackage{multirow}
\usepackage{placeins}
\usepackage{xcolor}
\usepackage{xurl}
\usepackage{hyperref}

\makeatletter
\renewcommand\@cite[2]{\textsuperscript{#1\if@tempswa , #2\fi}}
\renewcommand\@biblabel[1]{#1.}
\makeatother

\AddToHook{env/table/begin}{\setlength{\belowcaptionskip}{6pt}}

\newcommand{\tablebodyformat}{%
  \small
  \setlength{\tabcolsep}{3pt}%
  \renewcommand{\arraystretch}{1.20}%
}

\makeatletter
\newcommand{\backmattersection}{%
  \@startsection{section}{1}{\z@}%
  {-2.0ex \@plus -0.5ex \@minus -0.2ex}%
  {1.0ex \@plus 0.2ex}%
  {\normalfont\Large\bfseries}}
\makeatother

\title{DrivAerRim: A large-scale full-vehicle CFD\\
dataset for wheel rim aerodynamics}

\author{
Chao Xia$^{1,*}$,
Xi Zeng$^{2}$,
Alexey Vdovin$^{1}$,
Huanxia Wei$^{2}$,\\
Simone Sebben$^{1}$,
Carl Lewerth$^{1}$,
Qing Jia$^{2}$,
Zhigang Yang$^{2}$\\[0.5em]
$^{1}$Department of Mechanical Engineering, Chalmers University of\\
Technology, Gothenburg, 412 96, Sweden\\
$^{2}$College of Automotive and Energy Engineering, Tongji University,\\
Shanghai, 201804, China\\
$^{*}$Corresponding author: Chao Xia (chao.xia@chalmers.se)
}

\date{}

\begin{document}

\maketitle

\begin{abstract}
The wheel region contributes substantially to vehicle aerodynamic drag, and rim geometry strongly influences the flow in this region. However, publicly available automotive aerodynamics datasets offer limited variation in rim design. We present DrivAerRim, a large-scale dataset of full-vehicle computational fluid dynamics simulations covering 904 DeepWheel rim designs with diverse spoke layouts, opening areas and topologies. All cases use the same DrivAer Estateback, operating conditions and numerical setup, with rim geometry as the sole design variable. Wheel rotation is modelled using moving reference frame regions around the rims and rotating wall conditions on the deformed tyres. The numerical setup is assessed through mesh sensitivity analysis and comparisons with wind tunnel measurements. The resulting 9.32~TB release includes wheel geometries, vehicle and component force coefficients, accumulated drag, surface pressure, wall shear stress, and flow fields around the vehicle and wheels. Together, these data support surrogate model training and evaluation across diverse rim geometries, as well as aerodynamic optimisation and generative design to accelerate wheel rim development for vehicle drag reduction.
\end{abstract}

\section*{Background \& Summary}

Data-driven methods are increasingly used for aerodynamic prediction, optimisation and generative design. Evaluating large numbers of design variants using computational fluid dynamics (CFD) simulations or wind tunnel tests is costly. Publicly available CFD datasets support the development and benchmarking of these methods across a range of fluid mechanics applications.
AeroFlowData\cite{aeroflowdata} covers nearly 40 configurations and more than 500 experimental and numerical flow conditions for hypersonic vehicles, civil aircraft, and turbomachinery blades at high Reynolds numbers. HiLiftAeroML\cite{hiliftaeroml} comprises 1,800 wall-modelled large-eddy simulations (WMLES) for NASA CRM-HL configurations. BlendedNet\cite{blendednet} contains 8,830 Reynolds-averaged Navier--Stokes (RANS) cases spanning geometric and flight-condition variations in blended wing body aircraft. SuperWing\cite{superwing} covers 4,239 parameterised transonic wings through 28,856 RANS solutions and includes surface and volumetric flow fields for surrogate model development. Luminary SHIFT\cite{shift-suv} similarly combines CFD datasets for specific applications with pretrained artificial intelligence (AI) models.

Within automotive aerodynamics, public datasets have progressed from simplified canonical bodies to full-vehicle representations. AhmedML\cite{ahmedml} and WindsorML\cite{windsorml} feature high-fidelity simulations of simplified squareback bluff bodies. More recent datasets based on DrivAer model move closer to full-vehicle applications. DrivAerNet\cite{drivaernet} comprises 4,000 RANS simulations of parametric fastback configurations with wheels, mirrors, and detailed underbody modelling. DrivAerNet++\cite{drivaernetpp} extends this setting to 8,000 RANS cases across fastback, notchback, and estateback configurations with various underbody and wheel options. DrivAerML\cite{drivaerml} consists of 500 high-fidelity cases generated using delayed detached eddy simulation (DDES) for a parametrically morphed DrivAer notchback. DrivAerStar\cite{drivaerstar} provides 12,000 RANS simulations with vehicle components representative of industrial applications, cooling flow, and free-form deformation parameters. SHIFT-SUV\cite{shift-suv} contributes a growing collection of DDES simulations for parametrically morphed AeroSUV estateback and fastback variants. These resources also support geometric deep learning surrogate models for predicting aerodynamic force coefficients and flow fields from vehicle geometry. The methods range from point-cloud\cite{pointnet} and mesh-based graph networks\cite{meshgraphnets} to neural operators and transformers, including DoMINO\cite{domino}, GAOT\cite{gaot}, Transolver\cite{transolver}, GeoTransolver\cite{geotransolver}, AB-UPT\cite{ab-upt}, and SMART\cite{smart}. Table~\ref{tab:automotive-datasets} compares these datasets with this work.

\begin{table}[htbp]
\centering
\caption{Comparison of existing automotive CFD datasets.}
\label{tab:automotive-datasets}
\begingroup
\tablebodyformat
\begin{tabular}{>{\raggedright\arraybackslash}p{0.23\linewidth}>{\raggedright\arraybackslash}p{0.245\linewidth}>{\centering\arraybackslash}p{0.09\linewidth}>{\centering\arraybackslash}p{0.125\linewidth}>{\raggedright\arraybackslash}p{0.235\linewidth}}
\hline
Dataset/source & Geometry & Cases & \mbox{CFD method} & Wheel treatment \\
\hline
AhmedML\cite{ahmedml} & Ahmed body & 500 & DDES & No wheels \\
WindsorML\cite{windsorml} & Windsor body & 355 & WMLES & No wheels \\
DrivAerNet\cite{drivaernet} & DrivAer fastback & 4,000 & RANS & Rotating wall \\
DrivAerNet++\cite{drivaernetpp} & DrivAer, three rear ends & 8,000 & RANS & Rotating wall \\
DrivAerML\cite{drivaerml} & DrivAer notchback & 500 & DDES & Stationary wheels \\
DrivAerStar\cite{drivaerstar} & DrivAer, three rear ends & 12,000 & RANS & Rotating wall \\
SHIFT-SUV\cite{shift-suv} & AeroSUV, two rear ends & 1,000+ & DDES & Rotating wall \\
\hline
\mbox{DrivAerRim (this work)} & DrivAer Estateback & 904 & RANS & Rotating wall + MRF \\
\hline
\end{tabular}
\endgroup
\end{table}

Despite this progress, most public automotive aerodynamics datasets and surrogate benchmarks focus on shape variation at the full-vehicle scale. Their geometric variants are generated mainly through parametric morphing of overall vehicle dimensions, while wheel variation is limited to a small set of parameters, including rim closure, wheel width and wheel diameter. As a result, these datasets provide limited coverage of detailed rim geometries. Most datasets that model wheel rotation also use rotating wall boundary conditions alone (Table~\ref{tab:automotive-datasets}).
These limitations restrict the range of wheel geometries and rotation treatments available for surrogate training. A component-wise analysis in CarBench identifies the wheel region as particularly challenging for prediction\cite{carbench}. Wheel rotation alters the flow close to the ground, and fine rim features generate localized structures that are difficult for surrogate models to reproduce.

Accurate prediction in this region matters because wheels and wheelhouses account for approximately 25--35\% of the aerodynamic resistance of passenger vehicles, depending on vehicle configuration\cite{vdovin2013,vdovin2014,brandt2019,altmann2022}. Rotating wheels also generate ventilation resistance through pressure and shear forces on the tyres, rims and brake surfaces\cite{vdovin2013,vdovin2014}. Changes in rim coverage, outer radial features, spoke layout and rim depth affect both drag and the flow leaving the wheelhouse\cite{brandt2019}. Drag differences between rim designs may amount to only a few counts (one count corresponds to 0.001). Predicted rim rankings are sensitive to tyre deformation and to whether wheel rotation is modelled using a rotating wall, moving reference frame (MRF), or rigid body motion\cite{altmann2022}.
Existing rim studies establish the aerodynamic importance of rim geometry, but typically examine a limited set of production wheels or low-dimensional parametric families. Such parameterisations cover only a subset of the spoke layouts, opening patterns and other features found in contemporary rims, while generative design continues to expand this design space. Datasets constructed from the same parameterisations expose surrogate models to a restricted range of shapes and limit aerodynamic assessment across diverse rim geometries.

To address these gaps, we present DrivAerRim, a large-scale CFD dataset comprising RANS simulations of all 904 rim geometries from DeepWheel \cite{deepwheel} on a shared DrivAer Estateback model\cite{drivaer}. DeepWheel combines generative AI, topology optimisation and three-dimensional (3D) reconstruction to produce spoke layouts and topologies beyond a single parametric rim family. Wheel rotation is represented by MRF regions around the rims and rotating wall conditions on deformed tyres with longitudinal grooves and ground contact patches. The numerical setup is assessed through mesh sensitivity analysis and comparisons with wind tunnel measurements. DrivAerRim provides vehicle and component force coefficients, accumulated drag, surface pressure, wall shear stress, and flow fields around the vehicle and wheels. By linking diverse rim geometries to full-vehicle aerodynamic responses, the dataset enables systematic assessment of rim designs. It provides a basis for geometry deep learning surrogate model development and data-driven design exploration aimed at reducing vehicle drag.

\section*{Methods}

\subsection*{Dataset design}

To build a dataset for rim aerodynamics, we use the same DrivAer generic model for all cases. In each case, we assemble one DeepWheel rim design at all four wheel positions. The vehicle body, closed-cooling configuration, smooth underbody, deformed Rain tyres and wheel positions are unchanged across all cases. All cases use the same open-road domain and boundary conditions. The numerical setup also remains unchanged, including the meshing strategy, RANS settings and data-export definitions. The cylindrical MRF regions are constructed using a common strategy for all rim designs. Using a consistent vehicle configuration and simulation setup across all cases isolates the aerodynamic effects of rim geometry.

\subsection*{Vehicle geometry}

\begin{figure}[htbp]
\centering
\includegraphics[width=0.95\linewidth]{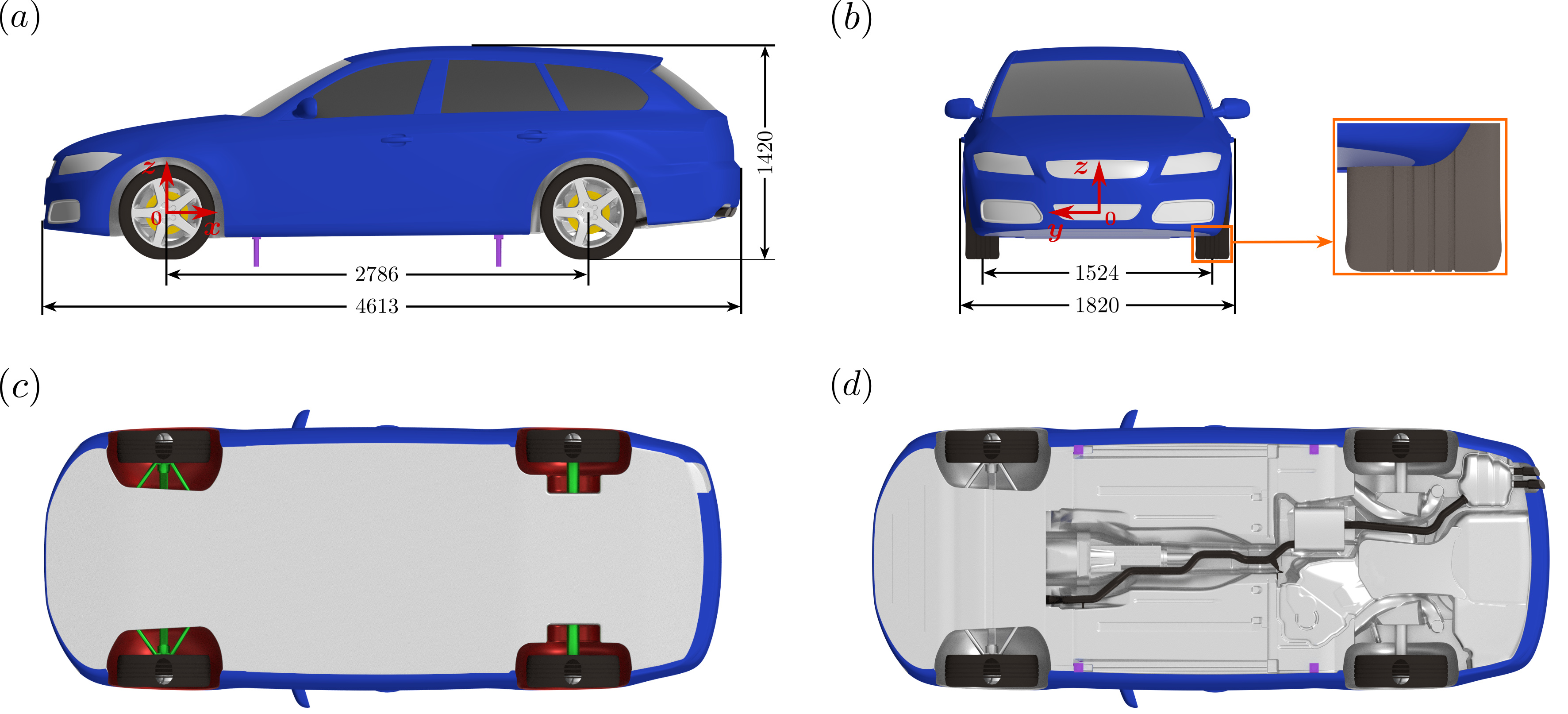}
\caption{Full-scale DrivAer Estateback geometry and component colours. (a) Side view with dimensions, wheel components and the purple restraint posts used only for validation. (b) Front view and Rain tyre tread. (c) Smooth underbody used for dataset generation, with \texttt{WheelHouse} (red) and \texttt{WheelSupport} (green). (d) Detailed underbody and restraint posts used for validation. All dimensions are in mm.}
\label{fig:vehicle-geometry}
\end{figure}

We use the full-scale closed-cooling DrivAer Estateback model as the reference vehicle (Fig.~\ref{fig:vehicle-geometry}). The vehicle is 4613 mm long, 1820 mm wide and 1420 mm high, with a wheelbase of 2786 mm and a track width of 1524 mm (Fig.~\ref{fig:vehicle-geometry}a,b). The DrivAer generic vehicle was introduced by Heft et al. and has since been investigated extensively in wind tunnel experiments and numerical studies\cite{drivaer,strangfeld2013,avadiar2018,ekman2020}. Later versions added an engine bay and cooling flow path\cite{wittmeier2015}, including the full-scale open-cooling model described by Hupertz et al.\cite{hupertz2018}. The wheel and brake installation follows the configurations used by Josefsson et al. in their studies of DrivAer wheel aerodynamics\cite{josefsson2022wind,josefsson2022tyre,josefsson2024reynolds}.

For dataset generation, we use the smooth underbody shown in Fig.~\ref{fig:vehicle-geometry}c and omit the wind tunnel restraint posts. The smooth underbody represents the relatively flat and enclosed underfloor layouts common in electric vehicles. For validation, we use the detailed underbody and the purple rocker panel restraint posts shown in Fig.~\ref{fig:vehicle-geometry}a,d to match the model tested in the Volvo Cars Aerodynamic Wind Tunnel. We use the Rain (R) tread pattern investigated by Josefsson et al.\cite{josefsson2022tyre} (Fig.~\ref{fig:vehicle-geometry}b). The tyre geometry includes longitudinal grooves and static deformations, including sidewall bulging near the ground contact patches. These details are retained because wheel flow is sensitive to the tyre shoulder shape and contact patch geometry.

Fig.~\ref{fig:vehicle-geometry} defines the component names used throughout the dataset. The black tyre, grey rim and yellow brake system together form the \texttt{Wheel} assembly. In Fig.~\ref{fig:vehicle-geometry}c, the red wheelhouse is named \texttt{WheelHouse}, while the green suspension geometry is named \texttt{WheelSupport}. The purple restraint posts are used only in the validation geometry. When a wheel position is specified, the suffixes \texttt{FL}, \texttt{RL}, \texttt{FR} and \texttt{RR} denote front left, rear left, front right and rear right, respectively.

\subsection*{Rim geometry}

\begin{figure}[htb]
\centering
\includegraphics[width=0.95\linewidth]{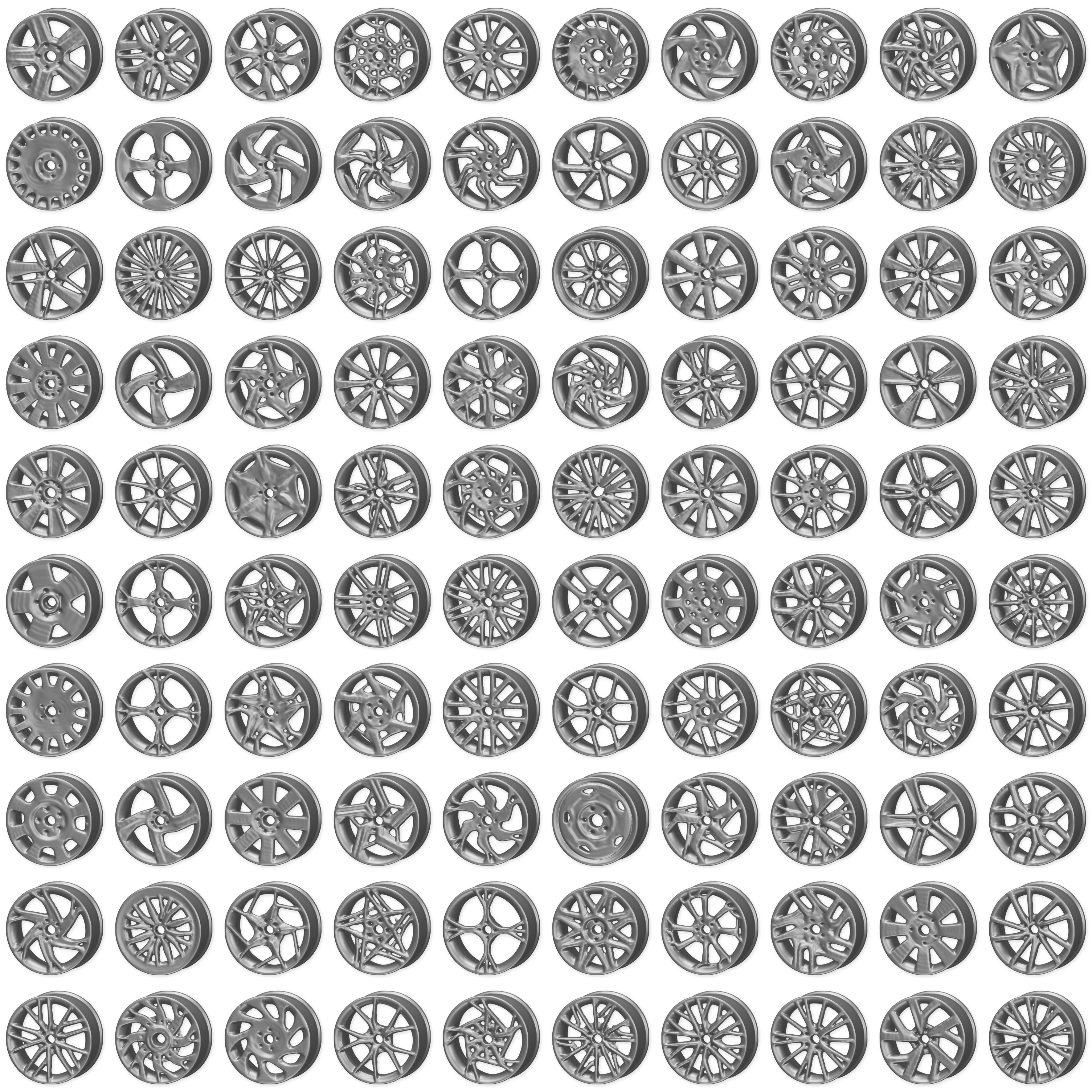}
\caption{Examples of rim geometries used in this study.}
\label{fig:rim-geometry}
\end{figure}

We used all 904 rim geometries from the DeepWheel dataset\cite{deepwheel}. DeepWheel contains synthetic three-dimensional rim designs produced using a workflow that combines generative design, topology optimisation, three-dimensional reconstruction and structural-performance evaluation. Examples are shown in Fig.~\ref{fig:rim-geometry}. The collection spans a wide range of spoke counts and shapes, opening areas and overall topologies. Before meshing, we repaired each source geometry to remove open edges, duplicate facets, non-manifold features and local intersections that could interfere with surface wrapping or volume meshing. The repaired rims were then scaled where necessary and aligned with the local coordinate system at each wheel position.

\subsection*{Numerical method}\label{sec:numerical}

Three-dimensional, steady incompressible RANS simulations were performed with the finite-volume solver Simcenter STAR-CCM+ (Siemens, Version 2506). All dataset cases used the realizable \(k-\epsilon\) (RKE) two-layer model with two-layer all \(y^+\) wall treatment. To select the turbulence model, RKE, the lag elliptic blending (LagEB) \(k-\epsilon\) model and the shear stress transport (SST) \(k-\omega\) model were compared with wind tunnel data. An SST variant with a realizability coefficient of 1.2 and \(a_1=1\), labelled SST\_a1 in the validation figures, was also evaluated. The drag difference and surface pressure comparisons reported in the section Technical Validation supported the choice of RKE for dataset generation.

Solver settings followed the Siemens external aerodynamics best practice guidelines\cite{siemens2506bpg}. The coupled implicit flow solver used second-order spatial discretization and Roe flux-difference splitting (Roe FDS) without blending. Convection in the turbulence equations also used second-order discretization. Automatic Courant--Friedrichs--Lewy (CFL) control adjusted the pseudo time step, with the initial, minimum and maximum CFL values at their defaults. Constant explicit relaxation and expert initialization with default grid sequencing parameters were used. Algebraic multigrid (AMG) used Gauss--Seidel smoothing for both equation systems. The coupled flow solver used a V-cycle with stabilized biconjugate gradient (BiCGStab) acceleration, while the turbulence solver used a flexible cycle without acceleration. Both AMG solvers used a maximum of 30 cycles and a convergence tolerance of 0.1 per linear solve.

Each dataset case was run for 3000 solver iterations. The final 500 iterations were used to calculate the reported means. The runtime was approximately 0.5~h per case on 2,880 CPU cores, corresponding to about 1,440 CPU core-hours. Across all dataset cases, the estimated total was approximately \(1.30\times10^6\) CPU core-hours.

\subsubsection*{Computational domain}

\begin{figure}[htbp]
\centering
\includegraphics[width=0.95\linewidth]{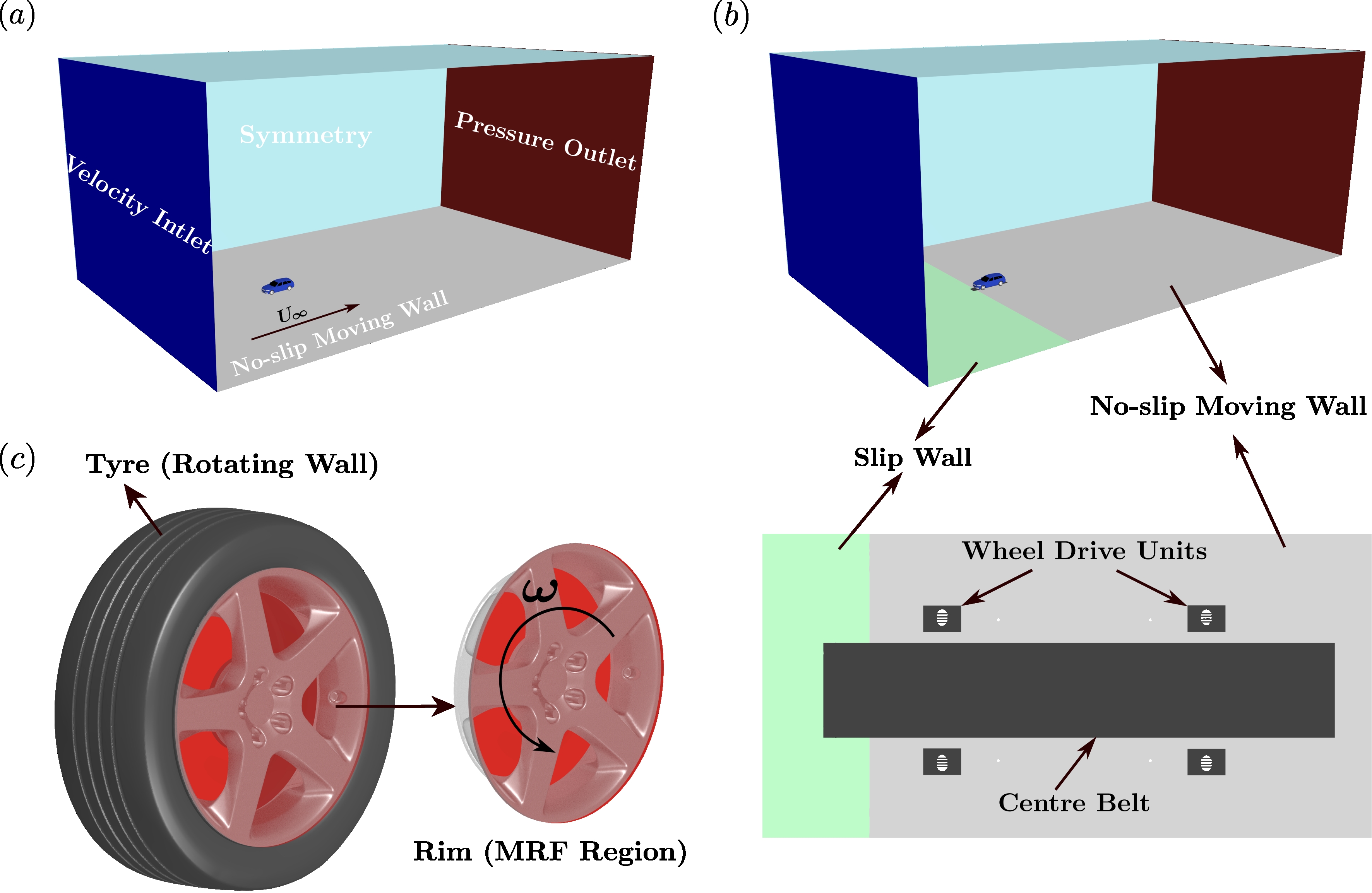}
\caption{Computational configurations. (a) Open-road domain for dataset generation. (b) Wind tunnel ground configuration for validation. (c) Rotating wall boundary conditions on the tyres and cylindrical MRF regions around the rims.}
\label{fig:computational-domain}
\end{figure}

All dataset cases used the open-road computational domain shown in Fig.~\ref{fig:computational-domain}a, following the setup for DrivAer wheel aerodynamics\cite{josefsson2022wind}. The rectangular domain extended 15.2 vehicle lengths in the streamwise direction, 8.7 vehicle lengths in width and 6.6 vehicle lengths in height. The inlet and outlet were located 4.3 and 9.9 vehicle lengths upstream and downstream of the vehicle, respectively. All cases were simulated at zero yaw. At the inlet, a uniform velocity of \(U_\infty=140~\mathrm{km/h}\) and a turbulence intensity of 0.1\% were prescribed. The downstream boundary was specified as a pressure outlet, while symmetry conditions were imposed on the top and lateral boundaries. The ground was modelled as a no-slip moving wall with velocity \(U_\infty\).

For numerical validation, the open-road ground treatment was replaced with a configuration representative of the Volvo Cars Aerodynamic Wind Tunnel (Fig.~\ref{fig:computational-domain}b). The upstream ground section, shown in light green, was specified as a slip wall to represent the effect of the wind tunnel boundary layer suction system on the approaching flow. The centre belt and four belts beneath the wheel-drive units formed a five-belt system, with each belt modelled as a no-slip moving wall. All other boundary conditions matched those used for the dataset cases. This configuration was used only for mesh sensitivity and validation.

\subsubsection*{Wheel rotation modelling\label{sec:rotation}}

Wheel rotation was represented using MRF regions around the rims and rotating wall boundary conditions on the tyres (Fig.~\ref{fig:computational-domain}c), following the Siemens guidelines\cite{siemens2506bpg}. Within each MRF region, the governing equations were solved in a frame rotating at a prescribed angular velocity. A separate rotating reference frame was assigned to each of the four rim regions. For each frame, the rotation axis passed through the corresponding wheel centre and was aligned with the wheel axle, and the angular velocity \(\omega\) was specified independently. Using only a rotating wall condition for the rims can impose zero tangential velocity on some spoke side faces\cite{altmann2022}. Tangential velocity was prescribed on the tyre surfaces, while the deformed tyre shape, longitudinal grooves and contact-patch geometry were retained.

The Siemens guidelines\cite{siemens2506bpg} recommend MRF regions confined to the spoke passages, but also allow cylindrical regions with outer interfaces close to the rim and tyre surfaces. Because passage-confined regions could not be generated automatically for every rim geometry, we used cylindrical MRF regions for all cases. The gap between the outer MRF interface and the outer rim face was less than 5~mm in every case. The tyre sidewalls shielded these compact regions.

\subsubsection*{Aerodynamic coefficients and reference quantities}

Full-vehicle aerodynamic forces were expressed using the drag, lift and side force coefficients:
\begin{equation}
C_D=\frac{F_x}{0.5\rho U_\infty^2 A_{\mathrm{ref}}},\qquad
C_L=\frac{F_z}{0.5\rho U_\infty^2 A_{\mathrm{ref}}},\qquad
C_S=\frac{F_y}{0.5\rho U_\infty^2 A_{\mathrm{ref}}},
\label{eq:force-coefficients}
\end{equation}
where \(F_x\), \(F_z\) and \(F_y\) denote the full-vehicle drag, lift and side force in the \(x\), \(z\) and \(y\) directions, respectively. The air density was \(\rho=1.204~\mathrm{kg/m^3}\), and the reference area was \(A_{\mathrm{ref}}=2.16~\mathrm{m^2}\).

The front and rear axle lift coefficients were obtained from the full-vehicle lift and pitching moment. Let \(M_{y,R}\) and \(M_{y,F}\) denote the \(+y\) pitching moments about reference points at ground height and the rear and front axle longitudinal positions, respectively. Using the wheelbase \(b=2.786~\mathrm{m}\) as the moment reference length gives
\begin{equation}
C_{LF}=\frac{M_{y,R}}{0.5\rho U_\infty^2 A_{\mathrm{ref}}b},\qquad
C_{LR}=-\frac{M_{y,F}}{0.5\rho U_\infty^2 A_{\mathrm{ref}}b},\qquad
C_L=C_{LF}+C_{LR}.
\label{eq:axle-lift-coefficients}
\end{equation}
The skin-friction coefficient was defined from the magnitude of the wall shear stress vector as
\begin{equation}
C_f=\frac{\left|\vec{\tau}_w\right|}{0.5\rho U_\infty^2},
\label{eq:skin-friction-coefficient}
\end{equation}
where \(\vec{\tau}_w=(\tau_x,\tau_y,\tau_z)\) contains the three Cartesian components of the wall shear stress.

The surface pressure coefficient was defined as
\begin{equation}
C_p=\frac{p-p_\infty}{0.5\rho U_\infty^2},
\label{eq:pressure-coefficient}
\end{equation}
where \(p\) is the local static pressure and \(p_\infty\) is the freestream static pressure.

The total pressure coefficient was defined using the same reference quantities as
\begin{equation}
C_{p,\mathrm{tot}}=\frac{p_{\mathrm{tot}}-p_\infty}{0.5\rho U_\infty^2},\qquad
p_{\mathrm{tot}}=p+0.5\rho\left|\mathbf{U}\right|^2,
\label{eq:total-pressure-coefficient}
\end{equation}
where \(\mathbf{U}=(U_x,U_y,U_z)\) is the velocity vector in the laboratory frame.

The ventilation drag coefficient expresses the aerodynamic resistance to wheel rotation as an equivalent drag coefficient. For each wheel, the resisting moment divided by the rolling radius gives the equivalent drag force. Following Vdovin et al.\cite{vdovin2014}, the ventilation drag coefficient of each wheel was calculated from its aerodynamic moment about the wheel centre as
\begin{equation}
C_{D,\mathrm{V},i}=\frac{M_{y,i}^{\mathrm{W}}}{0.5\rho U_\infty^2 A_{\mathrm{ref}}r_i},\qquad
r_i=z_{c,i}-z_{\mathrm{ground}}.
\label{eq:ventilation-drag}
\end{equation}
Here, \(M_{y,i}^{\mathrm{W}}\) is the \(+y\) moment of the pressure and shear forces on the wall surfaces of wheel \(i\). The reference radius \(r_i\) is the wheel centre height \(z_{c,i}\) above the ground level \(z_{\mathrm{ground}}\).

\subsubsection*{Mesh strategy}

\begin{figure}[H]
\centering
\includegraphics[width=0.85\linewidth]{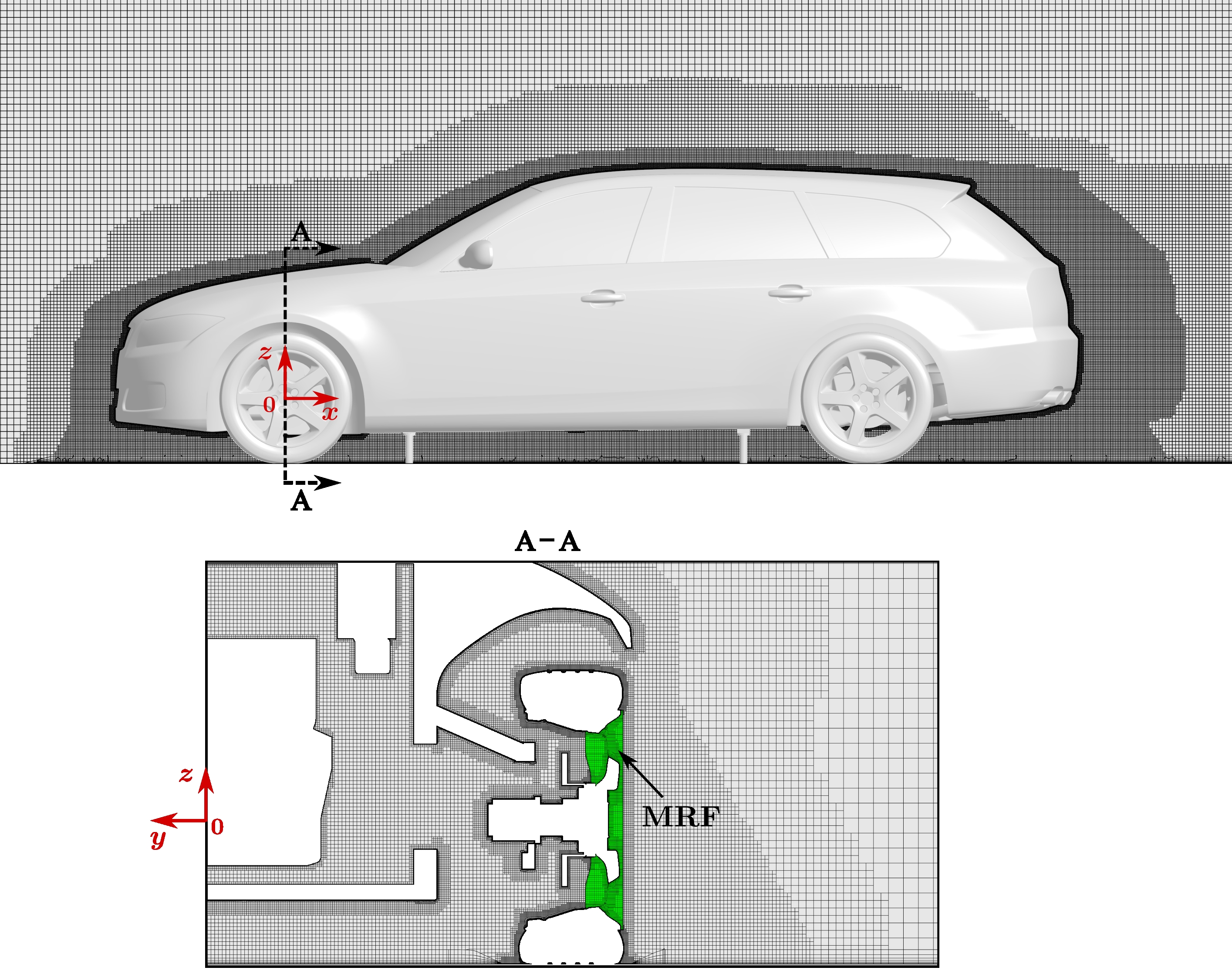}
\caption{Validation mesh with the wheelhouse and MRF region shown in section \(A\)-\(A\).}
\label{fig:mesh}
\end{figure}

A trimmed mesh composed predominantly of hexahedral cells was generated, with prism layers on all surfaces subject to no-slip conditions. Figure~\ref{fig:mesh} shows the baseline mesh for the validation geometry, including its detailed underbody and restraint posts. The same meshing controls were used for dataset generation with the smooth underbody. On the external vehicle surfaces, the target surface size was 1--4 mm. The wheel and rim regions used additional local refinement, with a target surface size of 1--2 mm around the tyres, rim openings, wheelhouses, tyre contact patches, and MRF volumes specific to each case. Ten prism layers were applied on surfaces with no-slip conditions, with a total thickness of 6 mm and a first-layer height of 0.015 mm, giving \(y^+<1\) on all surfaces.

\begin{figure}[H]
\centering
\includegraphics[width=0.80\linewidth]{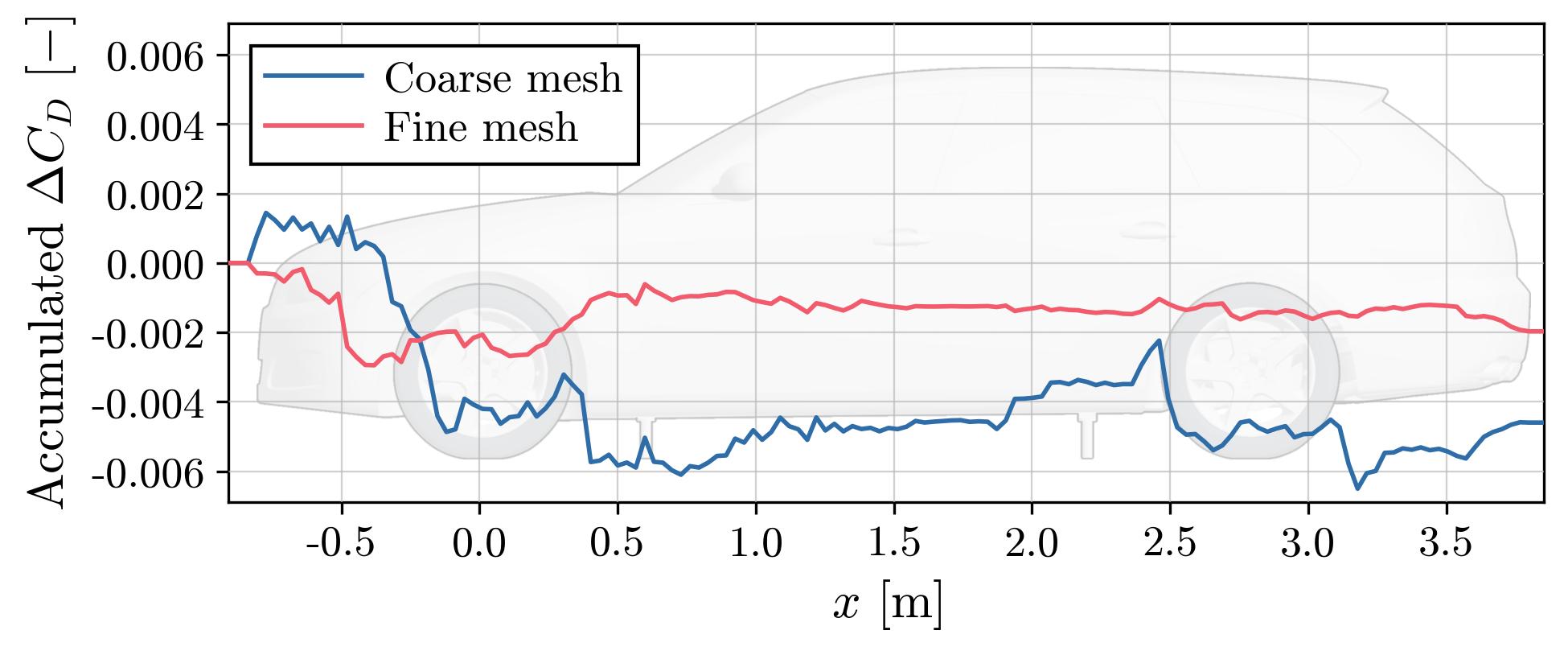}
\caption{Accumulated drag differences for the coarse and fine meshes relative to a baseline.}
\label{fig:mesh-sensitivity}
\end{figure}

For the mesh-sensitivity study, three refinement levels were generated by scaling the baseline target sizes while keeping all other meshing parameters unchanged. The coarse, baseline and fine meshes contained 45.32, 97.88 and 180.43 million cells, respectively. The accumulated drag difference was then evaluated relative to the baseline mesh,
\begin{equation}
\Delta C_D^{\mathrm{mesh}}(x)=C_D^{\mathrm{mesh}}(x)-C_D^{\mathrm{baseline}}(x),
\label{eq:mesh-drag-difference}
\end{equation}
so that positive or negative values indicate the streamwise accumulated drag offset from the baseline mesh. As shown in Fig.~\ref{fig:mesh-sensitivity}, the baseline and fine meshes remain close in accumulated drag along the vehicle length, whereas the coarse mesh shows larger deviations. The full-vehicle \(C_D\) differences relative to the baseline were \(-4.6\) counts for the coarse mesh and \(-2.0\) counts for the fine mesh. The baseline and fine meshes were close in both accumulated drag and full-vehicle \(C_D\). We therefore selected the baseline mesh to limit the computational resources required for the 904 dataset cases.

\subsection*{Dataset pipeline}

\begin{figure}[htb]
\centering
\includegraphics[width=0.99\linewidth]{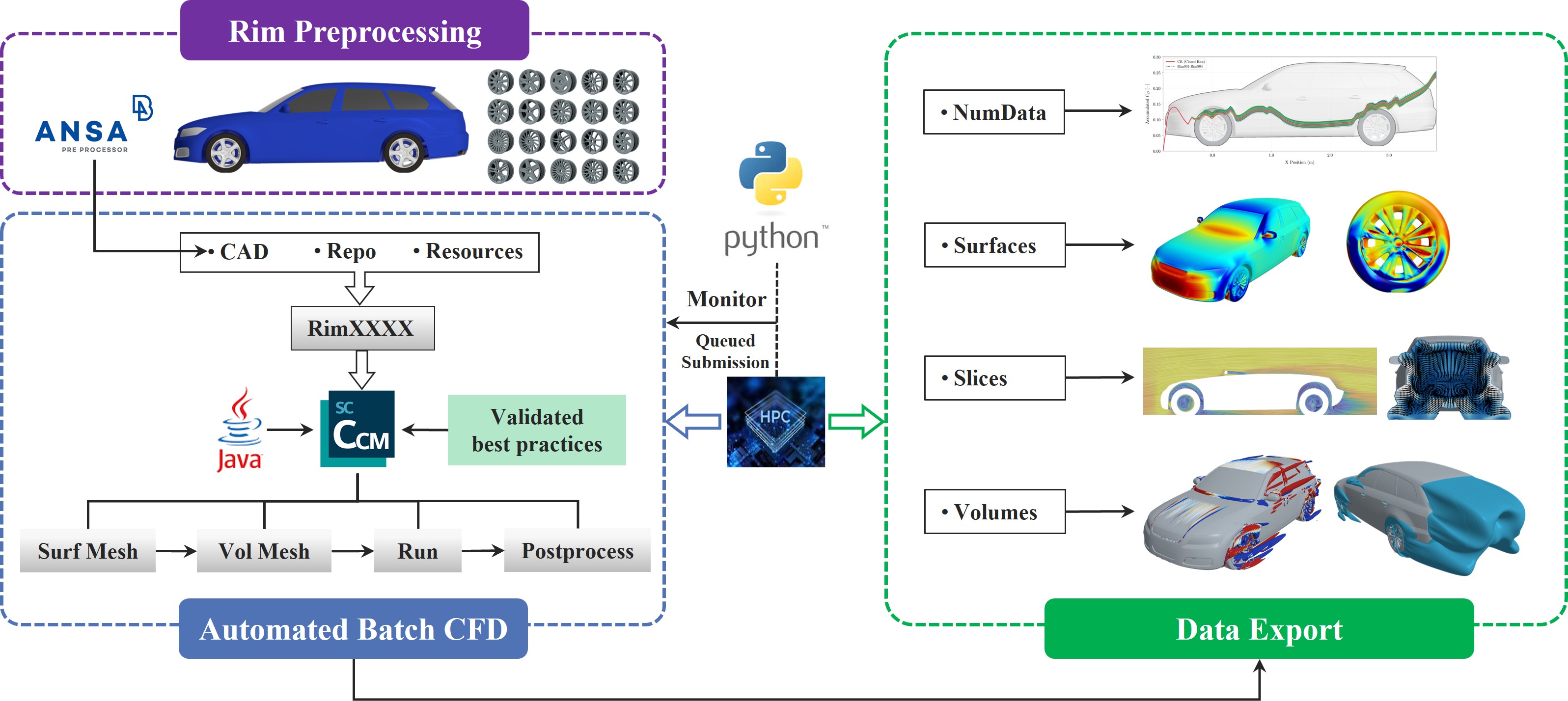}
\caption{DrivAerRim workflow.}
\label{fig:dataset-workflow}
\end{figure}

Dataset generation followed three stages: rim preprocessing, automated batch CFD and data export (Fig.~\ref{fig:dataset-workflow}). Each rim design defined one CFD case, with its simulation inputs and exported data organised in a \texttt{RimXXXX} directory (\texttt{Rim0001}--\texttt{Rim0904}).

\begin{itemize}
\setlength{\itemsep}{0.7em}
\setlength{\parsep}{0pt}
\setlength{\topsep}{0.4em}

\begin{samepage}
\item[\textbullet] \textbf{Rim Preprocessing.} ANSA was used to repair defects in the 904 rim STL surface meshes and to scale, align, and fit them to the common DrivAer wheel assembly. This produced consistently positioned, watertight rim geometries ready for simulation.
\end{samepage}

\item[\textbullet] \textbf{Automated Batch CFD.} 
The Python script \texttt{prepare.py} replicated the common case template to create 904 \texttt{RimXXXX} working directories. Each contained component geometry in \texttt{CAD} and Java macros in \texttt{repo} and \texttt{resources}. The script installed the preprocessed rim input in \texttt{CAD} for each case. The Python controller \texttt{batch\_run.py} submitted cases to the HPC scheduler with a limit on concurrently active jobs. It monitored job states and submitted the next pending case when a slot became available. Within each case, SLURM scripts invoked STAR-CCM+ Java macros for preparation, surface and volume meshing, solution, and postprocessing.

\item[\textbullet] \textbf{Data Export.} Postprocessing exported the numerical summaries, surface fields, planar slices, volumetric flow fields, and diagnostic images used in the released dataset. These outputs were written to the \texttt{NumData}, \texttt{Surfaces}, \texttt{Slices}, \texttt{Volumes}, and \texttt{Pictures} groups described in the Data Records section.

\end{itemize}

\FloatBarrier

\subsection*{Quality control during processing}\label{sec:QA}

The preprocessed rims were checked for watertightness and correct orientation before meshing. 
Rims that failed these checks were repaired.

Mesh quality checks focused on the constrained regions near the tyre contact patches. Automated monitors tracked the maximum velocity magnitude $|\mathbf{U}|_{\max}$ and turbulent viscosity ratio \(\mu_t/\mu\) during each calculation. When \(|\mathbf{U}|_{\max}>4U_\infty\) or \((\mu_t/\mu)_{\max}>10^4\), the cells responsible for these extrema were identified as problematic and removed before the solution continued. Such cells were found predominantly near the tyre contact patches. Mesh checks also verified the completion of prism layers, local refinement around the rims and contact patches, and overall cell quality.

\begin{figure}[htbp]
\centering
\includegraphics[width=0.95\linewidth]{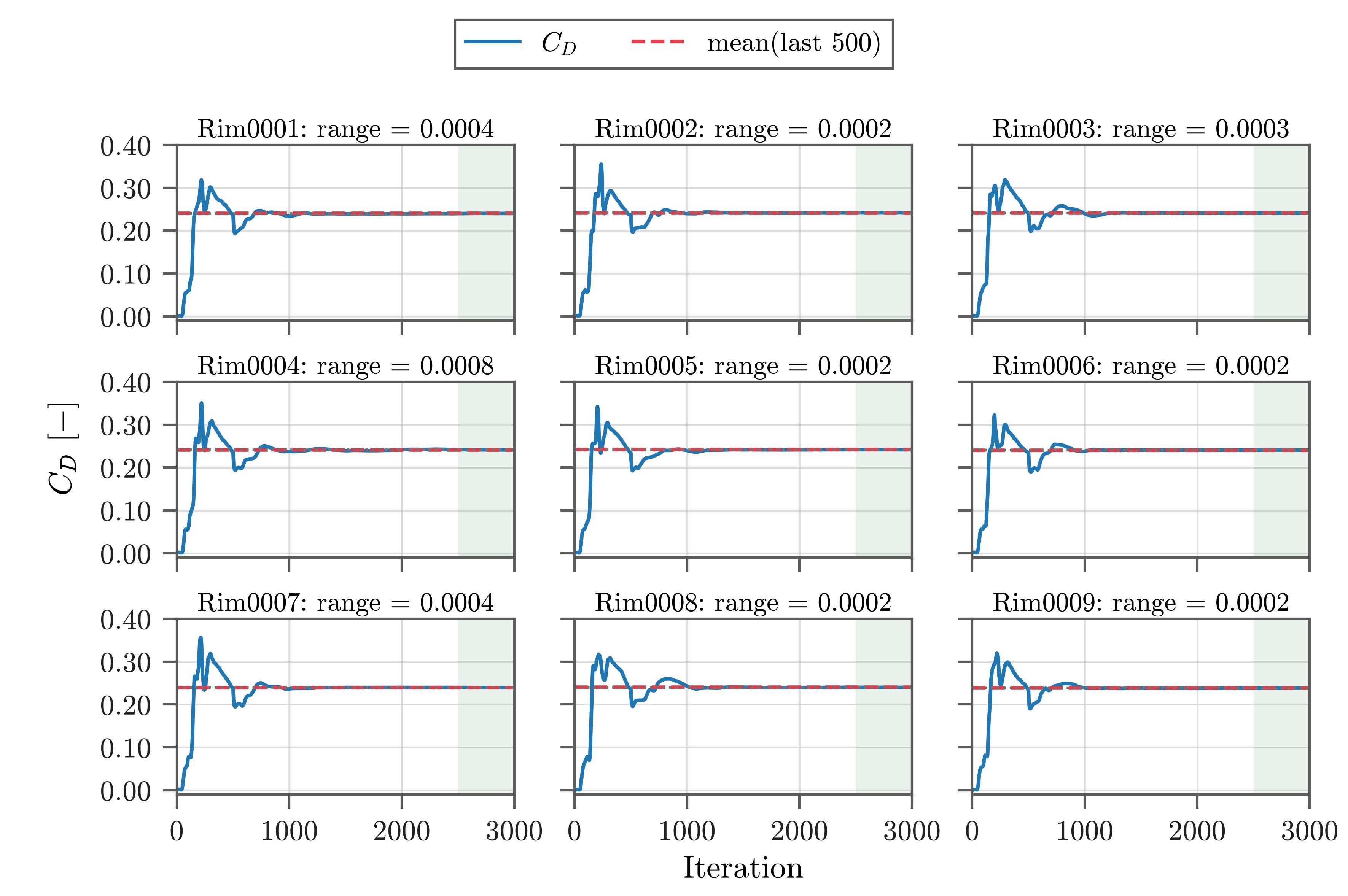}
\caption{Full-vehicle \(C_D\) histories for \texttt{Rim0001}--\texttt{Rim0009}. Shading and dashed lines indicate the final 500 iterations and their mean \(C_D\), respectively. Panel labels report the \(C_D\) range over these iterations.}
\label{fig:cd-convergence}
\end{figure}

Solver convergence was assessed over the final 500 iterations, which were also used to calculate the reported mean coefficients. For each case, the drag coefficient range \(R_{C_D}\) within this window was defined as
\begin{equation}
R_{C_D}=\max(C_D)-\min(C_D).
\label{eq:cd-convergence-range}
\end{equation}
The case was considered converged when \(R_{C_D}\leq0.003\). Figure~\ref{fig:cd-convergence} shows the full-vehicle \(C_D\) histories for the first nine cases. The same test was applied to all cases, together with inspection of residual histories and component force partitions. Cases showing an excessive range, abnormal force partitions, or failed field exports were flagged for review and rerun. Most failed runs were caused by defects in the rim surface meshes. The affected meshes were repaired, and the corresponding simulations were rerun. All dataset cases were completed and satisfied the convergence criterion.

\FloatBarrier

\section*{Data Records}

The DrivAerRim dataset is publicly available on Hugging Face at \url{https://doi.org/10.57967/hf/11461}\cite{drivaerrim-dataset}. The dataset version cited here corresponds to repository revision \texttt{b4831bd}. The repository contains approximately 9.32~TB of downloadable files in total. At the root level, the repository contains the \texttt{NumData} and \texttt{Geometry} directories and 904 \texttt{RimXXXX} case directories. The \texttt{NumData} directory contains aerodynamic coefficient and accumulated drag summaries for all cases, while \texttt{Geometry} contains STL geometries for the common vehicle body, wheel assemblies, wheelhouses, and wheel supports. Each \texttt{RimXXXX} directory groups the case-specific outputs for one rim configuration into four subdirectories: \texttt{Surfaces} for full-vehicle and wheel component surface data, \texttt{Slices} and \texttt{Volumes} for planar and volumetric flow fields, respectively, and \texttt{Pictures} for preview and convergence images. Table~\ref{tab:data-components} maps each repository location to its file pattern, size, and count, and Table~\ref{tab:field-variables} defines the output quantities provided in these files. The \texttt{Mean} prefix denotes an average over the final 500 solver iterations of the RANS calculation.
The contents of the data files in these directories are described below:

\begin{table}[!t]
\centering
\caption{DrivAerRim file organisation, sizes, and counts. \texttt{RimXXXX} entries are reported per case.}
\begingroup
\tablebodyformat
\setlength{\tabcolsep}{1.8pt}
\renewcommand{\arraystretch}{1.35}
\begin{tabular}{>{\centering\arraybackslash}m{0.122\linewidth}>{\centering\arraybackslash}m{0.167\linewidth}>{\hspace{0.5em}\raggedright\arraybackslash}m{0.453\linewidth}>{\centering\arraybackslash}m{0.10\linewidth}>{\centering\arraybackslash}m{0.115\linewidth}}
\hline
Root & Subdirectory & File & Size & No. of files \\
\hline
\multirow{2}{*}{\textbf{NumData}} & --- & \mbox{\texttt{Force\_Coefficients\_Combined.csv}} & 127 KiB & 1 \\
& --- & \mbox{\texttt{Cd\_Accumulated\_Combined.csv}} & 2 MiB & 1 \\
\hline
\multirow{7}{*}{\textbf{Geometry}} & \texttt{Car\_Body} & \mbox{\texttt{Car\_Body.stl}} & 276 MiB & 1 \\
& \texttt{Wheel\_FL} & \mbox{\texttt{Wheel\_FL\_XXXX.stl}} & 10 MiB & 904 \\
& \texttt{Wheel\_RL} & \mbox{\texttt{Wheel\_RL\_XXXX.stl}} & 10 MiB & 904 \\
& \texttt{WheelHouse\_FL} & \mbox{\texttt{WheelHouse\_FL.stl}} & 2.4 MiB & 1 \\
& \texttt{WheelHouse\_RL} & \mbox{\texttt{WheelHouse\_RL.stl}} & 2.4 MiB & 1 \\
& \texttt{WheelSupport\_FL} & \mbox{\texttt{WheelSupport\_FL.stl}} & 6.0 MiB & 1 \\
& \texttt{WheelSupport\_RL} & \mbox{\texttt{WheelSupport\_RL.stl}} & 2.9 MiB & 1 \\
\hline
\multirow{18}{*}{\textbf{RimXXXX}} & \multirow{7}{*}{\textbf{Surfaces}} & \mbox{\texttt{FullCar/FullCar\_XXXX.vtp}} & 275 MiB & \multirow{7}{*}{1 per case} \\
& & \mbox{\texttt{Wheel\_FL/Wheel\_FL\_XXXX.vtp}} & 37 MiB & \\
& & \mbox{\texttt{Wheel\_RL/Wheel\_RL\_XXXX.vtp}} & 37 MiB & \\
& & \mbox{\texttt{WheelHouse\_FL/WheelHouse\_FL\_XXXX.vtp}} & 3.7 MiB & \\
& & \mbox{\texttt{WheelHouse\_RL/WheelHouse\_RL\_XXXX.vtp}} & 3.6 MiB & \\
& & \mbox{\texttt{WheelSupport\_FL/WheelSupport\_FL\_XXXX.vtp}} & 2.4 MiB & \\
& & \mbox{\texttt{WheelSupport\_RL/WheelSupport\_RL\_XXXX.vtp}} & 1.2 MiB & \\
\cline{2-5}
& \multirow{3}{*}{\textbf{Slices}} & \mbox{\texttt{X/X\_<position>/X\_<position>\_XXXX.vtp}} & 184 MiB & 80 per case \\
& & \mbox{\texttt{Y/Y\_<position>/Y\_<position>\_XXXX.vtp}} & 170 MiB & 30 per case \\
& & \mbox{\texttt{Z/Z\_<position>/Z\_<position>\_XXXX.vtp}} & 179 MiB & 18 per case \\
\cline{2-5}
& \multirow{6}{*}{\textbf{Volumes}} & \mbox{\texttt{Wheel\_FL\_region/Wheel\_FL\_region\_XXXX.vtu}} & 134 MiB & \multirow{6}{*}{1 per case} \\
& & \mbox{\texttt{Wheel\_FL\_volume/Wheel\_FL\_volume\_XXXX.vtu}} & 983 MiB & \\
& & \mbox{\texttt{Wheel\_RL\_region/Wheel\_RL\_region\_XXXX.vtu}} & 132 MiB & \\
& & \mbox{\texttt{Wheel\_RL\_volume/Wheel\_RL\_volume\_XXXX.vtu}} & 952 MiB & \\
& & \mbox{\texttt{Wake\_volume/Wake\_volume\_XXXX.vtu}} & 597 MiB & \\
& & \mbox{\texttt{FullCar\_volume/FullCar\_volume\_XXXX.vtu}} & 5.9 GiB & \\
\cline{2-5}
& \multirow{2}{*}{\textbf{Pictures}} & \mbox{\texttt{<preview\_name>.webp}} & 25 MiB & 32 per case \\
& & \mbox{\texttt{Cd\_steady\_history.png}} & 94 KiB & 1 per case \\
\hline
\end{tabular}
\endgroup
\label{tab:data-components}
\end{table}

\begin{table}[!t]
\centering
\caption{Variables provided in the dataset files.}
\begingroup
\tablebodyformat
\renewcommand{\arraystretch}{1.45}
\begin{tabular}{>{\centering\arraybackslash}m{0.25\linewidth}>{\centering\arraybackslash}m{0.12\linewidth}>{\centering\arraybackslash}m{0.13\linewidth}>{\raggedright\arraybackslash}m{0.38\linewidth}}
\hline
\multicolumn{4}{c}{\textbf{\texttt{NumData/Force\_Coefficients\_Combined.csv}}} \\
Field or quantity & Units & Symbol & Description \\
\hline
\texttt{Cd} & \([{-}]\) & \(C_D\) & drag coefficient of the full vehicle \\
\texttt{Cl} & \([{-}]\) & \(C_L\) & total lift coefficient \\
\texttt{Clf} & \([{-}]\) & \(C_{LF}\) & front axle lift coefficient \\
\texttt{Clr} & \([{-}]\) & \(C_{LR}\) & rear axle lift coefficient \\
\texttt{Cs} & \([{-}]\) & \(C_S\) & side force coefficient \\
\texttt{Cd\_FL} & \([{-}]\) & \(C_{D,\mathrm{W,FL}}\) & front left wheel assembly drag coefficient \\
\texttt{Cd\_RL} & \([{-}]\) & \(C_{D,\mathrm{W,RL}}\) & rear left wheel assembly drag coefficient \\
\texttt{Cd\_vent\_FL} & \([{-}]\) & \(C_{D,\mathrm{V,FL}}\) & front left ventilation drag coefficient \\
\texttt{Cd\_vent\_RL} & \([{-}]\) & \(C_{D,\mathrm{V,RL}}\) & rear left ventilation drag coefficient \\
\texttt{Cd\_wh\_FL} & \([{-}]\) & \(C_{D,\mathrm{WH,FL}}\) & front left wheelhouse drag coefficient \\
\texttt{Cd\_wh\_RL} & \([{-}]\) & \(C_{D,\mathrm{WH,RL}}\) & rear left wheelhouse drag coefficient \\
\texttt{Cd\_ws\_FL} & \([{-}]\) & \(C_{D,\mathrm{WS,FL}}\) & front left wheel support drag coefficient \\
\texttt{Cd\_ws\_RL} & \([{-}]\) & \(C_{D,\mathrm{WS,RL}}\) & rear left wheel support drag coefficient \\
\texttt{Cl\_FL} & \([{-}]\) & \(C_{L,\mathrm{W,FL}}\) & front left wheel lift coefficient \\
\texttt{Cl\_RL} & \([{-}]\) & \(C_{L,\mathrm{W,RL}}\) & rear left wheel lift coefficient \\
\texttt{Cl\_wh\_FL} & \([{-}]\) & \(C_{L,\mathrm{WH,FL}}\) & front left wheelhouse lift coefficient \\
\texttt{Cl\_wh\_RL} & \([{-}]\) & \(C_{L,\mathrm{WH,RL}}\) & rear left wheelhouse lift coefficient \\
\texttt{Cl\_ws\_FL} & \([{-}]\) & \(C_{L,\mathrm{WS,FL}}\) & front left wheel support lift coefficient \\
\texttt{Cl\_ws\_RL} & \([{-}]\) & \(C_{L,\mathrm{WS,RL}}\) & rear left wheel support lift coefficient \\
\hline
\multicolumn{4}{c}{\textbf{\texttt{Surfaces/<surface>/<surface>\_XXXX.vtp}}} \\
\hline
\texttt{MeanPressure} & \([\mathrm{Pa}]\) & \(\overline{p}\) & static pressure \\
\texttt{MeanWSS} & \([\mathrm{Pa}]\) & \(\overline{\tau}_i\) & wall shear stress vector \\
\hline
\multicolumn{4}{c}{\textbf{\texttt{Slices/<sliceNormal>/<position>/<sliceNormal>\_<position>\_XXXX.vtp}}} \\
\hline
\texttt{MeanVelocity} & \([\mathrm{m/s}]\) & \(\overline{U}_i\) & velocity vector \\
\texttt{MeanPressure} & \([\mathrm{Pa}]\) & \(\overline{p}\) & static pressure \\
\texttt{MeanTotalTKE} & \([\mathrm{m^2/s^2}]\) & \(\overline{k}\) & total turbulent kinetic energy \\
\hline
\multicolumn{4}{c}{\textbf{\texttt{Volumes/<volume>/<volume>\_XXXX.vtu}}} \\
\hline
\texttt{MeanVelocity} & \([\mathrm{m/s}]\) & \(\overline{U}_i\) & velocity vector \\
\texttt{MeanPressure} & \([\mathrm{Pa}]\) & \(\overline{p}\) & static pressure \\
\texttt{MeanTotalTKE} & \([\mathrm{m^2/s^2}]\) & \(\overline{k}\) & total turbulent kinetic energy \\
\hline
\end{tabular}
\endgroup
\label{tab:field-variables}
\end{table}

\subsection*{NumData}

This directory contains two CSV summaries for all dataset cases. For each rim geometry, the \texttt{Force\_Coefficients\_Combined.csv} file contains the full-vehicle coefficients \(C_D\), \(C_L\), \(C_{LF}\), \(C_{LR}\), and \(C_S\), together with drag and lift coefficients for wheel components. The individual variables are listed in Table~\ref{tab:field-variables}. For example, \texttt{Cd\_FL} and \texttt{Cd\_RL} denote the drag coefficients of the front left and rear left wheel assemblies, respectively. The component groups follow the nomenclature shown in Fig.~\ref{fig:vehicle-geometry}. The location suffixes \texttt{FL}, \texttt{RL}, \texttt{FR}, and \texttt{RR} denote front left, rear left, front right, and rear right, respectively. The qualifiers \texttt{wh}, \texttt{ws}, and \texttt{vent} denote wheelhouse, wheel support, and ventilation resistance, respectively. The \texttt{Cd\_Accumulated\_Combined.csv} file stores a streamwise trace of accumulated drag coefficient for each of the 904 rim geometries. This enables analysis of the vehicle regions in which different rim configurations alter the overall drag.

\subsection*{Geometry}

The \texttt{Car\_Body} subdirectory contains the common vehicle body STL mesh used for all cases. The \texttt{Wheel\_FL} and \texttt{Wheel\_RL} subdirectories contain the STL surface meshes for the front left and rear left wheel assemblies, respectively, with one mesh for each of the 904 rim geometries. The \texttt{WheelHouse\_FL}, \texttt{WheelHouse\_RL}, \texttt{WheelSupport\_FL}, and \texttt{WheelSupport\_RL} subdirectories each contain one STL mesh for the corresponding wheelhouse or wheel support component. These four components are identical in every rim case and are therefore distributed only once. At zero yaw, the front right and rear right wheels, wheelhouses, and wheel supports mirror the corresponding components on the left. Detailed component geometries and local surface and volume data are therefore provided only for the front left and rear left positions.

\subsection*{RimXXXX case directories}

Each case directory contains full-vehicle and component surface data, slice and volume flow fields, and preview images for one rim configuration. Surface and slice records are Visualization Toolkit (VTK) PolyData files (\texttt{.vtp}), whereas volumetric fields are VTK Unstructured Grid files (\texttt{.vtu}).

\begin{figure}[H]
\centering
\includegraphics[width=0.98\linewidth]{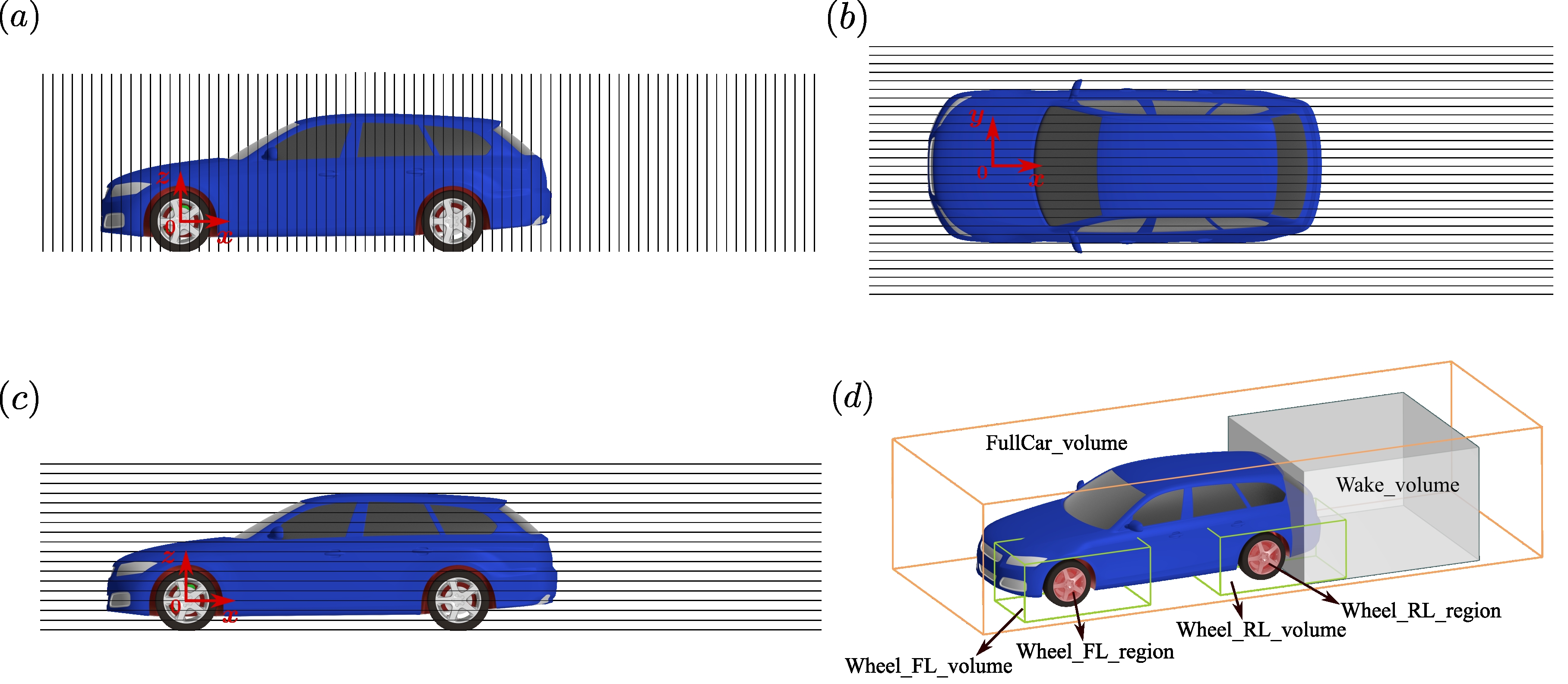}
\caption{Exported spatial data. (a) Slices normal to the \(x\) axis. (b) Slices normal to the \(y\) axis. (c) Slices normal to the \(z\) axis. (d) MRF regions and extraction volumes around the wheels, in the wake, and around the full vehicle.}
\label{fig:slices-volume}
\end{figure}

\begin{itemize}
\setlength{\itemsep}{0.45em}
\setlength{\parsep}{0pt}
\setlength{\topsep}{0.4em}

\item[\textbullet] \textbf{Surfaces.} Seven surface groups are provided for each case (Table~\ref{tab:data-components}). The \texttt{FullCar} group represents the full-vehicle surface and contains approximately 4.53 million surface cells. The \texttt{Wheel}, \texttt{WheelHouse}, and \texttt{WheelSupport} prefixes identify the wheel assembly, wheelhouse, and wheel support surfaces, respectively. The suffixes \texttt{FL} and \texttt{RL} identify the front left and rear left locations. Each wheel assembly record contains approximately 600,000 surface cells, compared with 60,000 for each wheelhouse and 30,000 for each wheel support. All seven \texttt{.vtp} files contain the \texttt{MeanPressure} and \texttt{MeanWSS} arrays listed in Table~\ref{tab:field-variables}.

\item[\textbullet] \textbf{Slices.} The \texttt{Slices} directory contains data sampled on 128 planes for each case. The \texttt{X}, \texttt{Y}, and \texttt{Z} subdirectories identify the plane normal, whereas the nested position directories and file names identify the plane location in millimetres. As illustrated in Fig.~\ref{fig:slices-volume}a--c, the extraction layout comprises 80 planes normal to \(x\) from \(-1.4\) to 6.5 m, 30 planes normal to \(y\) from \(-1.5\) to 1.5 m, and 18 planes normal to \(z\) from \(-0.3\) to 1.4 m. The planes have a nominal spacing of 0.1 m. Each \texttt{.vtp} slice record contains the \texttt{MeanVelocity}, \texttt{MeanPressure}, and \texttt{MeanTotalTKE} arrays listed in Table~\ref{tab:field-variables}.

\item[\textbullet] \textbf{Volumes.} Six volumetric regions are exported for each case, as shown in Fig.~\ref{fig:slices-volume}d. The \path{Wheel_FL_region} and \path{Wheel_RL_region} records coincide with the front left and rear left MRF regions (red) around the rims and spokes (Fig.~\ref{fig:computational-domain}c). Each MRF region contains approximately 2 million volume cells. The \path{Wheel_FL_volume} and \path{Wheel_RL_volume} records define larger extraction boxes near the wheels (green outlines), each containing approximately 13 million volume cells. The front left box spans \(x=-0.8\) to 1.0 m, \(y=-1.25\) to \(-0.25\) m, and \(z=-0.3\) to 0.5 m. The rear left box spans \(x=2.0\) to 3.8 m, \(y=-1.25\) to \(-0.25\) m, and \(z=-0.3\) to 0.5 m. The \path{Wake_volume} region (grey) spans \(x=3.2\) to 5.7 m, \(y=-1.25\) to 1.25 m, and \(z=-0.3\) to 1.25 m. It contains approximately 8.5 million volume cells and supports analysis of how rim geometry affects the vehicle wake. The \path{FullCar_volume} record (orange outline) covers a fluid volume surrounding the full vehicle, complementing the local wheel and wake data. This region contains approximately 83.85 million volume cells per case. It spans \(x=-1.5\) to 6.5 m, \(y=-1.5\) to 1.5 m, and \(z=-0.3\) to 1.5 m. All six \texttt{.vtu} volume records contain the \texttt{MeanVelocity}, \texttt{MeanPressure}, and \texttt{MeanTotalTKE} arrays listed in Table~\ref{tab:field-variables}.

\item[\textbullet] \textbf{Pictures.} Each case includes 32 \texttt{.webp} preview images for rapid inspection and quality control. These show \texttt{MeanPressure} and \texttt{MeanWSS} surface fields for the front left and rear left wheels from multiple views. They also include selected slice flow fields, volumetric \(Q\)-criterion visualisations, and total pressure coefficient isosurfaces. Each case additionally contains \texttt{Cd\_steady\_history.png}, which shows the full-vehicle \(C_D\) history.
\end{itemize}

\FloatBarrier
\section*{Data Overview}

\begin{figure}[p]
\centering
\includegraphics[width=0.90\linewidth]{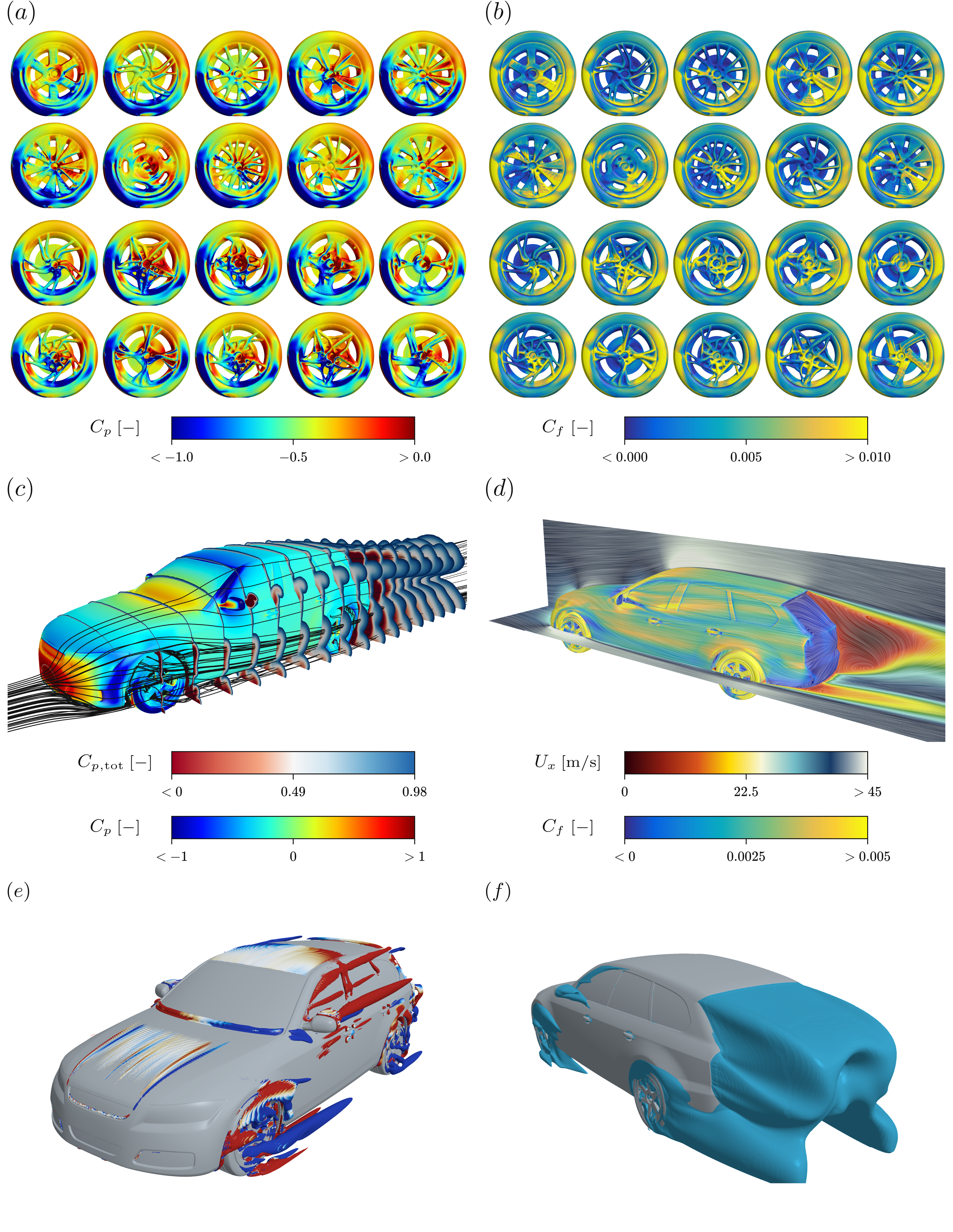}
\caption{Representative surface and flow data. (a) \(C_p\) and (b) \(C_f\) on the front left wheel for the 10 lowest \(C_D\) cases (top two rows) and 10 highest \(C_D\) cases (bottom two rows). (c--f) \texttt{Rim0265}: (c) \(C_p\), streamlines and \(C_{p,\mathrm{tot}}\) slices; (d) \(C_f\), wall shear stress oil flow patterns and \(U_x\) planes; (e) \(Q=2000~\mathrm{s}^{-2}\) isosurfaces coloured by \(\omega_x\) from \(-20\) to \(20~\mathrm{s}^{-1}\); and (f) \(C_{p,\mathrm{tot}}\) isosurface.}
\label{fig:surface-flow-overview}
\end{figure}

Across the 904 cases, the full-vehicle drag coefficient ranges from 0.2334 to 0.2551, with a mean of approximately 0.243. Figure~\ref{fig:surface-flow-overview} provides representative examples of the released surface data and flow fields. Panels (a,b) show \(C_p\) and \(C_f\) on the front left wheel for cases with low and high \(C_D\). Panels (c--f) show surface, slice and volume records for \texttt{Rim0265}. Corresponding records are available for every case in the directories described in Data Records.

\begin{samepage}
\section*{Technical Validation}

This section evaluates the numerical approach used to generate DrivAerRim. Four RANS models were compared with measurements from the Volvo Cars Aerodynamic Wind Tunnel using full-vehicle drag differences for two geometry modifications and surface pressure distributions at six vehicle sections\cite{josefsson2022wind}. The comparisons support the use of RKE for dataset generation.
\par
\end{samepage}

\subsection*{Drag differences}

\begin{figure}[!t]
\centering
\includegraphics[width=0.95\linewidth]{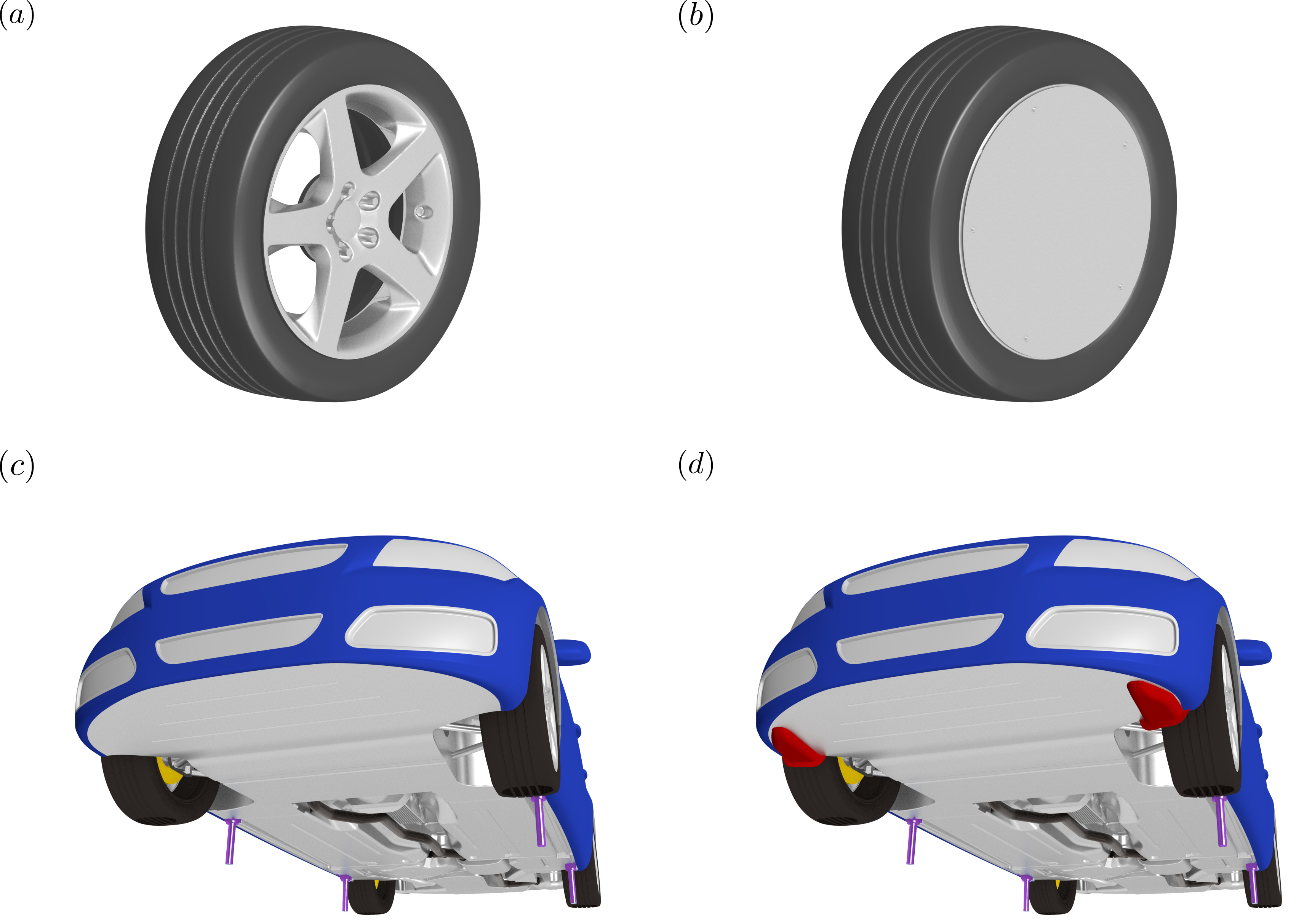}
\caption{Configurations for drag validation. (a) Open rim (OR). (b) Closed rim (CR). (c) Without front wheel deflectors (Wo\_FWD). (d) With front wheel deflectors (W\_FWD), highlighted in red.}
\label{fig:validation-configurations}
\end{figure}

Drag validation assessed whether the numerical setup reproduced changes in \(C_D\) caused by local geometry modifications around the wheels. The open rim (OR) and closed rim (CR) configurations (Fig.~\ref{fig:validation-configurations}a,b) evaluated the effect of rim closure. A second pair, without front wheel deflectors (Wo\_FWD) and with front wheel deflectors (W\_FWD), evaluated the effect of adding the deflectors (Fig.~\ref{fig:validation-configurations}c,d). The corresponding differences in full-vehicle drag coefficient were defined as
\begin{equation}
\Delta C_D^{\mathrm{OR}-\mathrm{CR}} =
C_D(\mathrm{OR}) - C_D(\mathrm{CR}),
\label{eq:rim-drag-difference}
\end{equation}
and
\begin{equation}
\Delta C_D^{\mathrm{Wo\_FWD}-\mathrm{W\_FWD}} =
C_D(\mathrm{Wo\_FWD}) - C_D(\mathrm{W\_FWD}).
\label{eq:deflector-drag-difference}
\end{equation}
Positive values indicate drag reductions due to rim closure or the addition of front wheel deflectors.

The validation compared four RANS models: RKE, LagEB, SST and SST\_a1. Experimental error bars were based on the wind tunnel test repeatability of \(\pm0.001\) in \(C_D\) for each configuration\cite{ljungskog2020}. The repeatability values for the two configurations were combined in quadrature without a covariance term, giving \(\pm\sqrt{2}\times0.001\) for each measured drag difference. Computational error bars represent the iterative variability of each predicted drag difference. For each configuration pair, \(R_1\) and \(R_2\) denote the \(C_D\) ranges defined in Eq.~(\ref{eq:cd-convergence-range}). Each predicted drag difference was reported as \(\Delta C_D \pm E_{\Delta C_D}\), where
\begin{equation}
E_{\Delta C_D}=\sqrt{R_1^2+R_2^2}.
\label{eq:drag-difference-variability}
\end{equation}

\begin{figure}[!htb]
\centering
\includegraphics[width=0.85\linewidth]{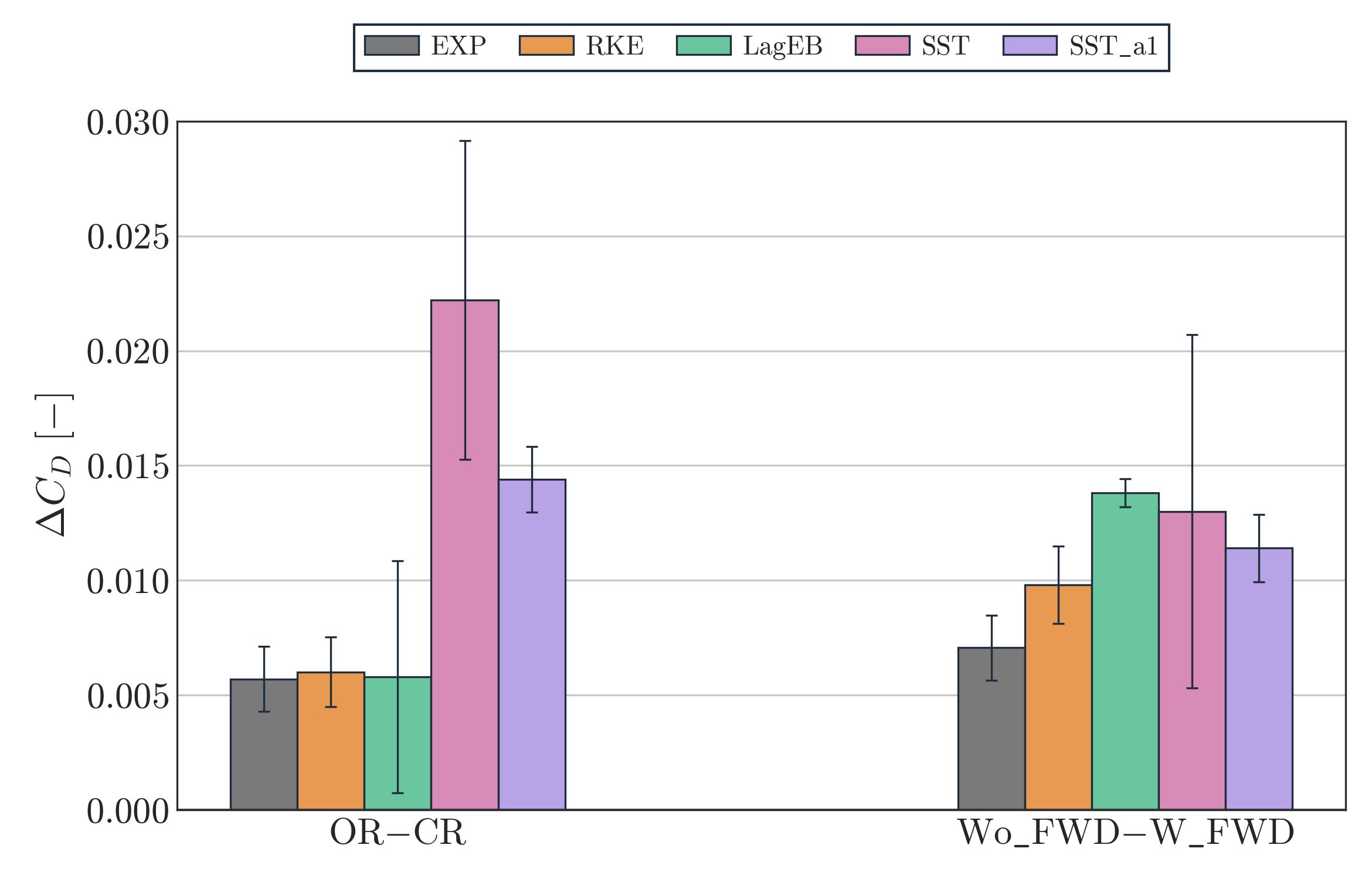}
\caption{Experimental measurements and CFD predictions of full-vehicle drag reductions caused by rim closure (\(\mathrm{OR}-\mathrm{CR}\)) and the addition of front wheel deflectors (\(\mathrm{Wo\_FWD}-\mathrm{W\_FWD}\)). Error bars indicate experimental repeatability and CFD iterative variability.}
\label{fig:validation-delta-cd}
\end{figure}

For rim closure, RKE and LagEB predicted \(\Delta C_D^{\mathrm{OR}-\mathrm{CR}}\) of approximately 0.006, in close agreement with the experimental value (Fig.~\ref{fig:validation-delta-cd}). SST and SST\_a1 predicted approximately 0.022 and 0.014, respectively. Even the lower SST\_a1 prediction was more than twice the experimental value. All four models captured the direction of the drag change caused by rim closure, but only RKE and LagEB reproduced its magnitude closely. The SST predictions varied most over the sampled iterations, while LagEB also showed substantial variation.

\enlargethispage{\baselineskip}
For the deflector comparison, all four models predicted a positive \(\Delta C_D^{\mathrm{Wo\_FWD}-\mathrm{W\_FWD}}\), consistent with the experimental result, but overestimated its magnitude. RKE predicted approximately 0.010 compared with 0.007 in the experiment, giving the smallest discrepancy. SST\_a1, SST and LagEB predicted approximately 0.011, 0.013 and 0.014, respectively. LagEB, which agreed closely for rim closure, gave the largest discrepancy for the deflector modification. The SST predictions again varied most over the sampled iterations, whereas RKE provided the most consistent drag difference predictions across both comparisons.

\subsection*{Surface pressure coefficients}

\begin{figure}[H]
\centering
\includegraphics[width=\linewidth]{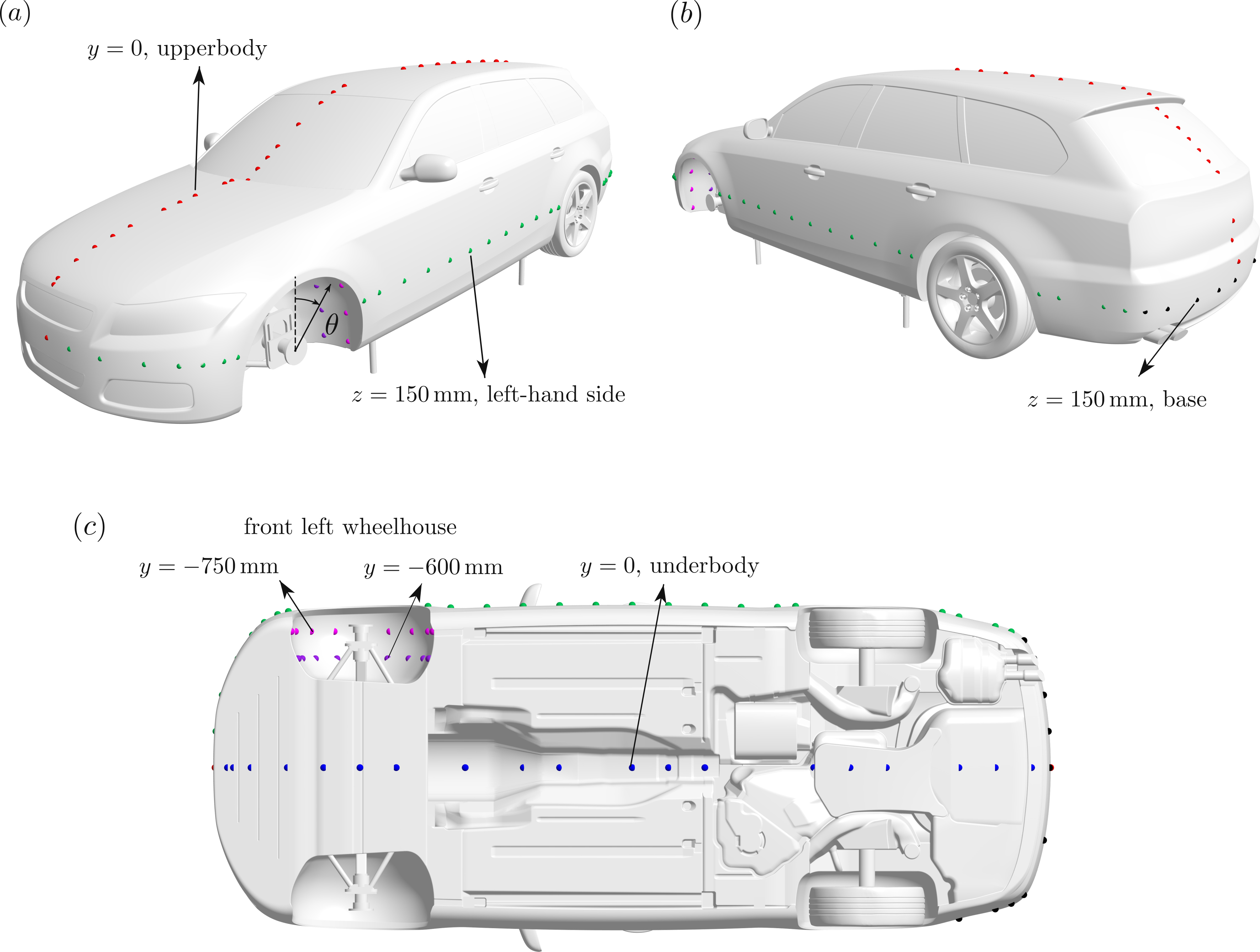}
\caption{Pressure tap layout on the DrivAer Estateback validation geometry. (a) Front left, (b) rear left and (c) underbody views. Red and blue markers denote the upperbody and underbody centreline taps. Green and black markers denote the left side and base taps at \(z=150\)~mm. Magenta and purple markers denote the wheelhouse rows at \(y=-750\) and \(-600\)~mm, respectively.}
\label{fig:pressure-probes}
\end{figure}

Figure~\ref{fig:pressure-probes} shows the six pressure tap groups used for validation on the DrivAer Estateback. Centreline taps were located at \(y=0\) on the upperbody and detailed underbody, while the left side and base taps followed the \(z=150\)~mm section. Two circumferential rows were placed in the front left wheelhouse at \(y=-750\) and \(-600\)~mm, with positions defined by the angle \(\theta\) about the wheel centre.

\begin{figure}[!htb]
\centering
\includegraphics[width=1.0\linewidth]{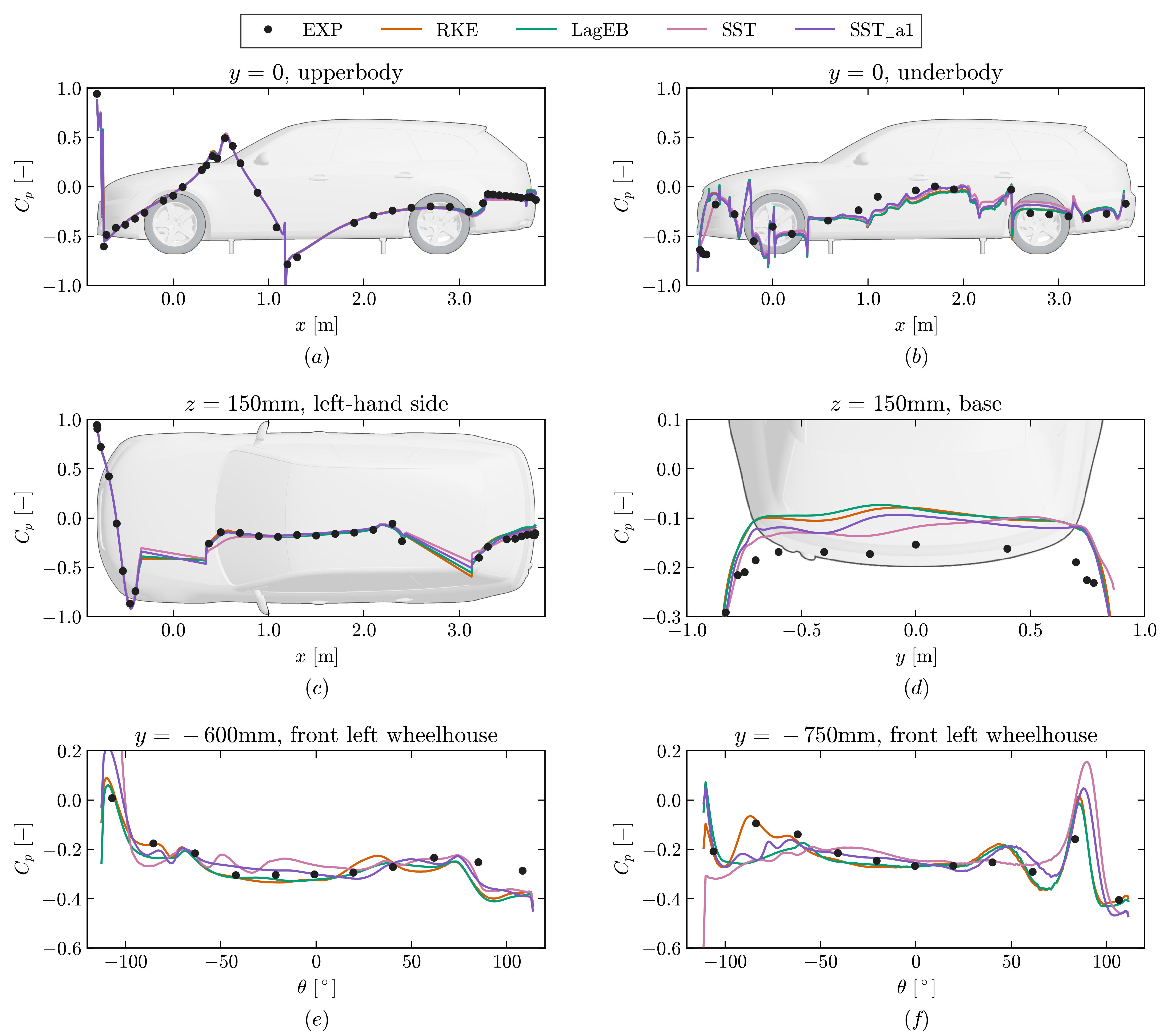}
\caption{Surface pressure validation for the open rim. (a) Upperbody. (b) Underbody. (c) Left side. (d) Base. (e,f) Front left wheelhouse.}
\label{fig:validation-cp}
\end{figure}

Figure~\ref{fig:validation-cp} compares the surface pressure coefficients predicted by the four RANS models with the wind tunnel measurements at six sections for the open rim configuration. The closed rim and front wheel deflector cases showed similar pressure trends and are not shown.
RKE reproduced the main measured pressure trends along the upperbody centreline and left side (Fig.~\ref{fig:validation-cp}a,c), and predictions were similar across the four RANS models. The predicted windscreen pressure rise, roof suction and subsequent recovery followed the experimental profiles. Differences were larger on the detailed underbody and base (Fig.~\ref{fig:validation-cp}b,d). RKE captured the broad underbody pressure recovery and, with LagEB, better matched measurements at \(x=2.6\) to 3.0~m. The SST variants were closer over part of the central floor and on the base, although all models underpredicted base suction.

In the wheelhouse at \(y=-600\)~mm (Fig.~\ref{fig:validation-cp}e), RKE and LagEB captured the broad negative pressure plateau. The SST variants overpredicted the positive peak near \(\theta=-110^\circ\), while SST\_a1 agreed better with some central measurements. All models overpredicted suction beyond \(\theta=80^\circ\). At \(y=-750\)~mm (Fig.~\ref{fig:validation-cp}f), RKE most closely reproduced the measured pressure rise near \(\theta=-90^\circ\). RKE and LagEB also gave closer predictions of the peak near \(\theta=80^\circ\) to \(90^\circ\) and the subsequent suction near \(\theta=110^\circ\) than the SST variants.

Together, the comparisons support RKE as a suitable choice for this dataset. Across the tested configurations, it combined the most consistent drag difference predictions with agreement for the main body and wheelhouse pressure trends, despite local pressure advantages of the SST variants.

\FloatBarrier

\section*{Usage Notes}

Given the 9.32~TB release size, users can retrieve numerical summaries with \texttt{NumData/**}, selected cases with \texttt{RimXXXX/**}, or individual files. Command-line and Python examples are available on the dataset repository page\cite{drivaerrim-dataset}.

\section*{Data Availability}

The DrivAerRim dataset is publicly available from Hugging Face at \url{https://doi.org/10.57967/hf/11461}\cite{drivaerrim-dataset} under the CC BY-NC 4.0 license. It contains the common geometry, aerodynamic coefficient and accumulated drag CSV files, and 904 case directories with surfaces, slices, volumes, and pictures. File formats and the directory structure are described in Data Records.

\section*{Code Availability}

Python scripts for case preparation, HPC job submission, job-state monitoring and aerodynamic-coefficient analysis, together with SLURM scripts for case execution and postprocessing, are available under CC BY-NC 4.0 license in the \texttt{Codes} directory at \url{https://github.com/Beyond0619/DrivAerRim/tree/main/Codes}. Setup instructions and dependencies are documented in the same directory.

\begin{samepage}
\backmattersection*{Author Contributions}
\setlength{\emergencystretch}{1em}

Chao Xia: Conceptualization, dataset generation, technical validation, and writing—original draft. Xi Zeng: Preprocessing of rim geometries and data post-processing. Alexey Vdovin: Writing—review and editing. Huanxia Wei: Data post-processing and writing—review and editing. Simone Sebben: Conceptualization and writing—review and editing. Carl Lewerth: Dataset generation. Qing Jia: Writing—review and editing. Zhigang Yang: Writing—review and editing.
\par
\end{samepage}

\backmattersection*{Competing Interests}

The authors declare no competing interests.

\backmattersection*{Acknowledgements}

The authors thank Ford of Europe, particularly Dr. Burkhard Hupertz and Manfred Lentzen, for providing access to the DrivAer and for sharing their expertise on the model. We also thank Erik Josefsson of Volvo Cars for supplying the tyre and rim data used for validation and for sharing the wind tunnel measurements, and Erik Sällström and Lukasz Sobieraj of Volvo Cars for their valuable technical discussions and advice throughout this work. We further thank Soyoung Yoo and Namwoo Kang for their helpful correspondence and clarification regarding the use of the DeepWheel dataset in this work.

\backmattersection*{Funding}

This work was supported by the Transport Area of Advance Seed Project 2026 at Chalmers University of Technology and the National Natural Science Foundation of China (Grant No. 52372360). Computational resources were provided by the National Academic Infrastructure for Supercomputing in Sweden (NAISS), funded by the Swedish Research Council.


\begin{thebibliography}{99}
\small
\setlength{\itemsep}{1.5pt}
\setlength{\parsep}{0pt}

\bibitem{aeroflowdata}
Zhang, W. et al. High-Reynolds-Number Turbulence Database: AeroFlowData. Sci. Data 12, 1500, \url{https://doi.org/10.1038/s41597-025-05846-4} (2025).

\bibitem{hiliftaeroml}
Ashton, N. et al. HiLiftAeroML: High-Fidelity Computational Fluid Dynamics Dataset for High-Lift Aircraft Aerodynamics. arXiv \url{https://doi.org/10.48550/arXiv.2605.19565} (2026).

\bibitem{blendednet}
Sung, N., Spreizer, S., Elrefaie, M., Samuel, K., Jones, M. C. \& Ahmed, F. BlendedNet: A Blended Wing Body Aircraft Dataset and Surrogate Model for Aerodynamic Predictions. In \textit{ASME 2025 International Design Engineering Technical Conferences and Computers and Information in Engineering Conference}, DETC2025-168977, \url{https://doi.org/10.1115/DETC2025-168977} (ASME, 2025).

\bibitem{superwing}
Yang, Y. et al. SuperWing: a comprehensive transonic wing dataset for data-driven aerodynamic design. Sci. Data 13, 1113, \url{https://doi.org/10.1038/s41597-026-07769-0} (2026).

\bibitem{shift-suv}
Luminary Cloud. SHIFT-SUV: High-Fidelity Computational Fluid Dynamics Dataset for SUV External Aerodynamics [Data set]. Hugging Face, \url{https://huggingface.co/datasets/luminary-shift/SUV} (2025).

\bibitem{ahmedml}
Ashton, N., Maddix, D. C., Gundry, S. \& Shabestari, P. M. AhmedML: High-Fidelity Computational Fluid Dynamics Dataset for Incompressible, Low-Speed Bluff Body Aerodynamics. arXiv \url{https://doi.org/10.48550/arXiv.2407.20801} (2024).

\bibitem{windsorml}
Ashton, N. et al. WindsorML: High-Fidelity Computational Fluid Dynamics Dataset For Automotive Aerodynamics. Advances in Neural Information Processing Systems 37, Datasets and Benchmarks Track, \url{https://proceedings.neurips.cc/paper_files/paper/2024/hash/42a59a5f35b1b3c3fd648397c88a7164-Abstract-Datasets_and_Benchmarks_Track.html} (2024).

\bibitem{drivaernet}
Elrefaie, M., Dai, A. \& Ahmed, F. DrivAerNet: A Parametric Car Dataset for Data-Driven Aerodynamic Design and Prediction. J. Mech. Des. 147, 041712, \url{https://doi.org/10.1115/1.4068104} (2025).

\bibitem{drivaernetpp}
Elrefaie, M., Morar, F., Dai, A. \& Ahmed, F. DrivAerNet++: A Large-Scale Multimodal Car Dataset with Computational Fluid Dynamics Simulations and Deep Learning Benchmarks. Advances in Neural Information Processing Systems 37, Datasets and Benchmarks Track, \url{https://doi.org/10.52202/079017-0016} (2024).

\bibitem{drivaerml}
Ashton, N. et al. DrivAerML: High-Fidelity Computational Fluid Dynamics Dataset for Road-Car External Aerodynamics. arXiv \url{https://doi.org/10.48550/arXiv.2408.11969} (2024).

\bibitem{drivaerstar}
Qiu, J. et al. DrivAerStar: An Industrial-Grade CFD Dataset for Vehicle Aerodynamic Optimization. Advances in Neural Information Processing Systems 38, Datasets and Benchmarks Track, \url{https://openreview.net/forum?id=GVB1M5K0ch} (2025).

\bibitem{pointnet}
Qi, C. R., Su, H., Mo, K. \& Guibas, L. J. PointNet: Deep Learning on Point Sets for 3D Classification and Segmentation. In \textit{Proceedings of the IEEE Conference on Computer Vision and Pattern Recognition} 652--660, \url{https://openaccess.thecvf.com/content_cvpr_2017/html/Qi_PointNet_Deep_Learning_CVPR_2017_paper.html} (2017).

\bibitem{meshgraphnets}
Pfaff, T., Fortunato, M., Sanchez-Gonzalez, A. \& Battaglia, P. W. Learning Mesh-Based Simulation with Graph Networks. In \textit{International Conference on Learning Representations}, \url{https://openreview.net/forum?id=roNqYL0_XP} (2021).

\bibitem{domino}
Ranade, R. et al. DoMINO: A Decomposable Multi-scale Iterative Neural Operator for Modeling Large Scale Engineering Simulations. arXiv \url{https://doi.org/10.48550/arXiv.2501.13350} (2025).

\bibitem{gaot}
Wen, S. et al. Geometry Aware Operator Transformer as an Efficient and Accurate Neural Surrogate for PDEs on Arbitrary Domains. Advances in Neural Information Processing Systems 38, \url{https://neurips.cc/virtual/2025/poster/118887} (2025).

\bibitem{transolver}
Wu, H., Luo, H., Wang, H., Wang, J. \& Long, M. Transolver: A Fast Transformer Solver for PDEs on General Geometries. In \textit{Proceedings of the 41st International Conference on Machine Learning}, PMLR 235, 53681--53705, \url{https://proceedings.mlr.press/v235/wu24r.html} (2024).

\bibitem{geotransolver}
Adams, C., Ranade, R., Cherukuri, R. \& Choudhry, S. GeoTransolver: Learning Physics on Irregular Domains Using Multi-scale Geometry Aware Physics Attention Transformer. arXiv \url{https://doi.org/10.48550/arXiv.2512.20399} (2025).

\bibitem{ab-upt}
Alkin, B. et al. AB-UPT: Scaling Neural CFD Surrogates for High-Fidelity Automotive Aerodynamics Simulations via Anchored-Branched Universal Physics Transformers. Trans. Mach. Learn. Res. \url{https://openreview.net/forum?id=nwQ8nitlTZ} (2025).

\bibitem{smart}
Hagnberger, J. \& Niepert, M. SMART: Scalable Mesh-free Aerodynamic Simulations from Raw Geometries using a Transformer-based Surrogate Model. In \textit{Proceedings of the 43rd International Conference on Machine Learning}, \url{https://openreview.net/forum?id=2dzU8sTOO3} (2026).

\bibitem{carbench}
Elrefaie, M., Shu, D., Klenk, M. \& Ahmed, F. CarBench: A Comprehensive Benchmark for Neural Surrogates on High-Fidelity 3D Car Aerodynamics. arXiv \url{https://doi.org/10.48550/arXiv.2512.07847} (2025).

\bibitem{vdovin2013}
Vdovin, A., Bonitz, S., Landstrom, C. \& Lofdahl, L. Investigation of Wheel Ventilation-Drag using a Modular Wheel Design Concept. SAE Int. J. Passeng. Cars - Mech. Syst. 6, 308--315, \url{https://doi.org/10.4271/2013-01-0953} (2013).

\bibitem{vdovin2014}
Vdovin, A., Lofdahl, L. \& Sebben, S. Investigation of Wheel Aerodynamic Resistance of Passenger Cars. SAE Int. J. Passeng. Cars - Mech. Syst. 7, 639--645, \url{https://doi.org/10.4271/2014-01-0606} (2014).

\bibitem{brandt2019}
Brandt, A., Berg, H., Bolzon, M. \& Josefsson, L. The Effects of Wheel Design on the Aerodynamic Drag of Passenger Vehicles. SAE Int. J. Adv. \& Curr. Prac. in Mobility 1, 1279--1299, \url{https://doi.org/10.4271/2019-01-0662} (2019).

\bibitem{altmann2022}
Altmann, P. et al. Experimental and Numerical Investigation of Rim Aerodynamics. SAE Technical Paper 2022-01-0891, \url{https://doi.org/10.4271/2022-01-0891} (2022).

\bibitem{deepwheel}
Yoo, S. \& Kang, N. DeepWheel: Generating a 3D Synthetic Wheel Dataset for Design and Performance Evaluation. J. Mech. Des. 148, 051702, \url{https://doi.org/10.1115/1.4069899} (2026).

\bibitem{drivaer}
Heft, A. I., Indinger, T. \& Adams, N. A. Introduction of a New Realistic Generic Car Model for Aerodynamic Investigations. SAE Technical Paper 2012-01-0168, \url{https://doi.org/10.4271/2012-01-0168} (2012).

\bibitem{strangfeld2013}
Strangfeld, C. et al. Experimental Study of Baseline Flow Characteristics for the Realistic Car Model DrivAer. SAE Technical Paper 2013-01-1251, \url{https://doi.org/10.4271/2013-01-1251} (2013).

\bibitem{avadiar2018}
Avadiar, T., Thompson, M. C., Sheridan, J. \& Burton, D. Characterisation of the Wake of the DrivAer Estate Vehicle. J. Wind Eng. Ind. Aerodyn. 177, 242--259, \url{https://doi.org/10.1016/j.jweia.2018.04.013} (2018).

\bibitem{ekman2020}
Ekman, P., Wieser, D., Virdung, T. \& Karlsson, M. Assessment of Hybrid RANS-LES Methods for Accurate Automotive Aerodynamic Simulations. J. Wind Eng. Ind. Aerodyn. 206, 104301, \url{https://doi.org/10.1016/j.jweia.2020.104301} (2020).

\bibitem{wittmeier2015}
Wittmeier, F. \& Kuthada, T. Open Grille DrivAer Model: First Results. SAE Int. J. Passeng. Cars Mech. Syst. 8, 252--260, \url{https://doi.org/10.4271/2015-01-1553} (2015).

\bibitem{hupertz2018}
Hupertz, B. et al. Introduction of a New Full-Scale Open Cooling Version of the DrivAer Generic Car Model. In \textit{Progress in Vehicle Aerodynamics and Thermal Management} 35--60, \url{https://doi.org/10.1007/978-3-319-67822-1_3} (Springer, 2018).

\bibitem{josefsson2022wind}
Josefsson, E., Hobeika, T. \& Sebben, S. Evaluation of wind tunnel interference on numerical prediction of wheel aerodynamics. J. Wind Eng. Ind. Aerodyn. 224, 104945, \url{https://doi.org/10.1016/j.jweia.2022.104945} (2022).

\bibitem{josefsson2022tyre}
Josefsson, E., Hobeika, T., Sebben, S. \& Urquhart, M. Investigation of Tyre Pattern Effect on the Aerodynamics of a Passenger Vehicle. J. Fluids Eng. 144, 111209, \url{https://doi.org/10.1115/1.4054851} (2022).

\bibitem{josefsson2024reynolds}
Josefsson, E., Urquhart, M. \& Sebben, S. Influence of rotating wheels on Reynolds number sensitivity of passenger vehicle drag. Phys. Fluids 36, 095182, \url{https://doi.org/10.1063/5.0226694} (2024).

\bibitem{siemens2506bpg}
Siemens Digital Industries Software. \textit{External Aerodynamics with Simcenter STAR-CCM+ Best Practice Guidelines (2506)}. Siemens Support Center, \url{https://support.sw.siemens.com/en-US/knowledge-base/KB000041270_EN_US}.

\bibitem{drivaerrim-dataset}
Xia, C. et al. DrivAerRim [Data set]. Hugging Face, \url{https://doi.org/10.57967/hf/11461} (2026).

\bibitem{ljungskog2020}
Ljungskog, E., Sebben, S. \& Broniewicz, A. Inclusion of the physical wind tunnel in vehicle CFD simulations for improved prediction quality. J. Wind Eng. Ind. Aerodyn. 197, 104055, \url{https://doi.org/10.1016/j.jweia.2019.104055} (2020).

\end{thebibliography}
\end{document}